\documentclass[pdflatex,sn-nature]{sn-jnl}

\usepackage{lmodern}
\usepackage{mathrsfs}
\usepackage{graphicx}%
\usepackage{multirow}%
\usepackage{amsmath,amssymb,amsfonts}%
\usepackage{amsthm}%
\usepackage[title]{appendix}%
\usepackage{xcolor}%
\usepackage{textcomp}%
\usepackage{manyfoot}%
\usepackage{booktabs}%
\usepackage{algorithm}%
\usepackage{algorithmicx}%
\usepackage{algpseudocode}%
\usepackage{listings}%
\usepackage{comment} 
\usepackage{bm}        
\usepackage{float}
\usepackage[normalem]{ulem}
\usepackage{soul}
\usepackage{placeins}

\renewcommand{\vec}{\mathbf}

\newcommand{\pwisein}{\left\{ \begin{array}{ll}}
\newcommand{\pwiseout}{\end{array}\right.}
\newcommand{\ket}[1]{\left| #1 \right\rangle}

\renewcommand{\eqref}[1]{Eq.~\ref{#1}}

\hypersetup{hypertexnames=false}
\begin{document}
\title{Boson Sampling as a Probe of Chaotic and Integrable Quantum Dynamics in a Photonic Chip}

\author[1,2]{\fnm{Yuancheng} \sur{Zhan}}\email{zhan0530@e.ntu.edu.sg}

\author[3]{\fnm{Khen} \sur{Cohen}}\email{khencohen@mail.tau.ac.il}
\equalcont{These authors contributed equally to this work.}

\author[2]{\fnm{Norman T.W.} \sur{Koo}} \email{normankoo@u.nus.edu}
\equalcont{These authors contributed equally to this work.}

\author[2,4]{\fnm{Kian Hwee} \sur{Lim} }\email{kianhwee\_lim@u.nus.edu}
\equalcont{These authors contributed equally to this work.}

\author[1,5]{\fnm{Hui} \sur{Zhang}}\email{jovie\_huizhang@tongji.edu.cn}

\author[1]{\fnm{Lingxiao} \sur{Wan}}\email{wanl0004@e.ntu.edu.sg}

\author[1]{\fnm{Sang Hoon} \sur{Chae}}\email{sanghoon.chae@ntu.edu.sg}

\author[6]{\fnm{Ai Qun} \sur{Liu}}\email{aiqun.liu@polyu.edu.hk}

\author[7]{\fnm{Victor M} \sur{Bastidas}}\email{victor.bastidas@ntt.com}

\author*[3]{\fnm{Yaron} \sur{Oz}}\email{yaronoz@tauex.tau.ac.il}

\author*[1,2,8]{\fnm{Leong-Chuan} \sur{Kwek}}\email{cqtklc@nus.edu.sg}

\affil[1]{\orgdiv{School of Electrical and Electronic Engineering (EEE)}, \orgname{Nanyang Technological University}, \orgaddress{\street{50 Nanyang Ave}, \city{Singapore}, \postcode{639798},  \country{Singapore}}}

\affil[2]{\orgdiv{Centre for Quantum Technologies (CQT)}, \orgname{National University of Singapore}, \orgaddress{\street{3 Science Drive 2}, \city{Singapore}, \postcode{117543}, \country{Singapore}}}

\affil[3]{\orgdiv{The School of Physics and Astronomy}, \orgname{Tel-Aviv University}, \orgaddress{\street{Haim Levanon}, \city{Tel Aviv}, \postcode{69978}, \country{Israel}}}

\affil[4]{\orgdiv{MajuLab}, \orgname{CNRS-UCA-SU-NUS-NTU International Joint Research Laboratory}, \orgaddress{\street{3 Science Drive 2}, \postcode{117543}, \country{Singapore}}}

\affil[5]{\orgdiv{School of Physics Science and Engineering}, \orgname{Tongji University}, \orgaddress{\street{Yangpu District}, \city{Shanghai}, \postcode{200092}, \country{China}}}

\affil[6]{\orgdiv{Institute of Quantum Technologies}, \orgname{The Hong Kong Polytechnic University}, \orgaddress{\street{11 Yuk Choi Road}, \city{Hong Kong}, \postcode{999077}, \country{China}}}

\affil[7]{\orgname{Basic Research Laboratories \& NTT Research Center for Theoretical Quantum Information, NTT, Inc.}, \orgaddress{\street{3-1 Morinosato-Wakamiya}, \city{Kanagawa}, \postcode{243-0198}, \country{Japan}}}

\affil[8]{\orgdiv{National Institute of Education}, \orgname{Nanyang Technological University}, \orgaddress{\street{1 Nanyang Walk}, \city{Singapore}, \postcode{637616}, \country{Singapore}}}

\abstract{
Quantum chaos plays a key role in understanding complex quantum dynamics, while integrated photonics offers unique advantages for quantum applications, including high-speed operation, scalability, and programmable unitary transformations. However, integrated photonic approaches to probing quantum chaos remain largely unexplored, owing to the absence of a clear connection between programmable photonic dynamics and established chaos diagnostics. In this work, we establish Fock-state boson sampling as a practical probe of quantum chaos by exploiting the sensitivity of multiphoton interference to the random-matrix properties of underlying single-particle unitary dynamics. More importantly, we design and fabricate a programmable quantum photonic chip to experimentally implement this framework, achieving the first integrated-photonic demonstration of quantum-chaos probes based on boson sampling.
Experimental results show that the three complementary probes proposed in this work, namely the distance to Porter--Thomas statistics, Shannon entropy, and Out-of-Time-Ordered-Correlator-equivalent observables, exhibit close agreement with theoretical predictions and consistently distinguish chaotic and integrable dynamics. Our work provides a scalable route for investigating complex quantum dynamics on programmable photonic platforms while leveraging the intrinsic advantages of boson sampling through multiphoton interference and complex output statistics.
}
\maketitle
\section{Introduction}
\label{sec: intro}
Quantum technologies have emerged as a central paradigm in modern science, underpinning advances in quantum information processing~\cite{ekert2002direct,nielsen2010quantum,bennett2000quantum}, cryptography~\cite{zhang2019integrated,gisin2002quantum,pirandola2020advances}, and quantum simulation~\cite{zhan2026loop,georgescu2014quantum}. Among the various physical platforms, photonic architectures offer distinct advantages for quantum applications~\cite{zhang2019integrated,luo2023recent,lenzini2018integrated,wang2020integrated,flamini2018photonic,zhang2026integrated}, including low decoherence, high-speed information processing, and the ability to implement large-scale, programmable unitary transformations via integrated optical networks~\cite{zhan2024physics,zhang2021optical,zhang2022resource}. These properties make integrated photonic systems an attractive platform for scalable quantum computing protocols and for experimentally studying fundamental aspects of quantum mechanics. Quantum chaos plays an important role in understanding thermalization, ergodicity, information spreading, and the emergence of complex dynamics in quantum systems~\cite{alessio2016chaos,chertkov2022holographic}. To characterize such behavior, quantum chaos is typically identified through spectral statistics, such as the spectral form factor (SFF)~\cite{cotler2017chaos}, or dynamically via the out-of-time-ordered correlators (OTOCs)~\cite{garcia2022out,maldacena2016a}. These signatures have been extensively investigated in trapped ions~\cite{garttner2017measuring}, superconducting circuits~\cite{xiao2021information}, and nuclear magnetic resonance~\cite{li2017measuring}. However, existing studies of quantum chaos often rely on specific dynamical models with direct semiclassical limits, such as atom--field interaction systems and related chaos models~\cite{Carlos_Santos_Hirsch_PRL_2019,yu2024neumann,lemos2012experimental}, making their implementation inherently model-dependent. Moreover, in optical systems, previous quantum chaos investigations have mainly focused on free-space implementations~\cite{yu2024neumann,Ying_PRA_2025,lemos2012experimental}. The implementation in programmable integrated photonic platforms remains largely unexplored because a clear connection between integrated photonics and probes of quantum chaos is still lacking. As a result, the advantages of integrated photonics have not yet been fully exploited in this area.

Boson sampling is a quantum computational model based on multiphoton interference in linear-optical networks, where output probabilities are determined by matrix permanents and are believed to be classically intractable to simulate for sufficiently large systems~\cite{madsen2022quantum,brod2019photonic,spring2013boson,zhu2025gbs}. Boson sampling has been extensively explored as a promising route toward quantum computational advantages in photonic systems, such as applications in molecular spectra~\cite{huh2015boson}, graph theory~\cite{deng_solving_graph_PRL_2023,huh2015boson}, and quantum machine learning~\cite{hoch2025quantum,wang2023experimental}. More importantly, the complexity of boson sampling output statistics suggests a potential connection to quantum chaos, where signatures of the underlying dynamics may be reflected in the measured probability distributions.

Building upon this potential connection, our first contribution is to establish Fock-state boson sampling as a practical probe of quantum chaos. We show that the temporal evolution of boson sampling output statistics carries signatures of the Hamiltonian dynamics and can therefore be used to distinguish chaotic and integrable behavior. This arises because the bosonic interference of indistinguishable photons makes the output probability distributions sensitive to the random-matrix properties of the underlying single-particle unitary evolution. As a result, signatures of chaotic dynamics can be extracted directly from the measured boson sampling statistics.
Based on this principle, we identify three complementary diagnostic probes from the temporal behavior of the boson sampling output distributions. First, we find that the distribution of output probability values approaches Porter--Thomas (PT) statistics~\cite{boxio2018characterizing} for chaotic dynamics at some predictable evolution time, where the underlying unitary evolution most closely approximates Haar-random behavior~\cite{cotler2017chaos}. Second, we show that the temporal evolution of the Shannon entropy~\cite{porter1956fluctuations,shannon1948mathematical} distinguishes the two regimes, where chaotic dynamics produce more delocalized output distributions and correspondingly larger entropy values than integrable dynamics. Third, we extract OTOC-equivalent observables from output probabilities through the mapping between boson sampling statistics and four-point OTOCs established in our work~\cite{bastidas2025equilibration}. Chaotic dynamics are found to exhibit stronger Hilbert-space delocalization and broader frequency content in their temporal evolution compared with integrable dynamics.

To experimentally implement this framework, our second contribution is the realization of quantum-chaos probes on a programmable integrated photonic platform, allowing the entire probing process to benefit from the scalability, reconfigurability, and quantum advantages of photonic chips. We employ a random-matrix framework in which a single parameter continuously tunes the crossover between integrable and chaotic regimes, providing a compact and naturally programmable implementation on photonic hardware. We design and fabricate a silicon integrated photonic chip capable of implementing programmable $M$-mode unitary transformations for boson sampling experiments with $N$ input photons. In our experimental demonstration, an eight-mode processor with two indistinguishable photons is programmed to realize unitary dynamics generated by integrable and chaotic random-matrix Hamiltonian ensembles. The resulting boson sampling statistics are obtained through multiphoton correlation measurements using superconducting nanowire single-photon detectors. Experimental results show that the proposed probes consistently distinguish integrable (Poissonian) and chaotic [Gaussian Orthogonal Ensemble (GOE)] dynamics, while providing, to the best of our knowledge, the first experimental realization of boson-sampling-based quantum-chaos probes in integrated photonic platforms.

\begin{figure*}[ht]
  \centering
  \includegraphics[width=1\linewidth]{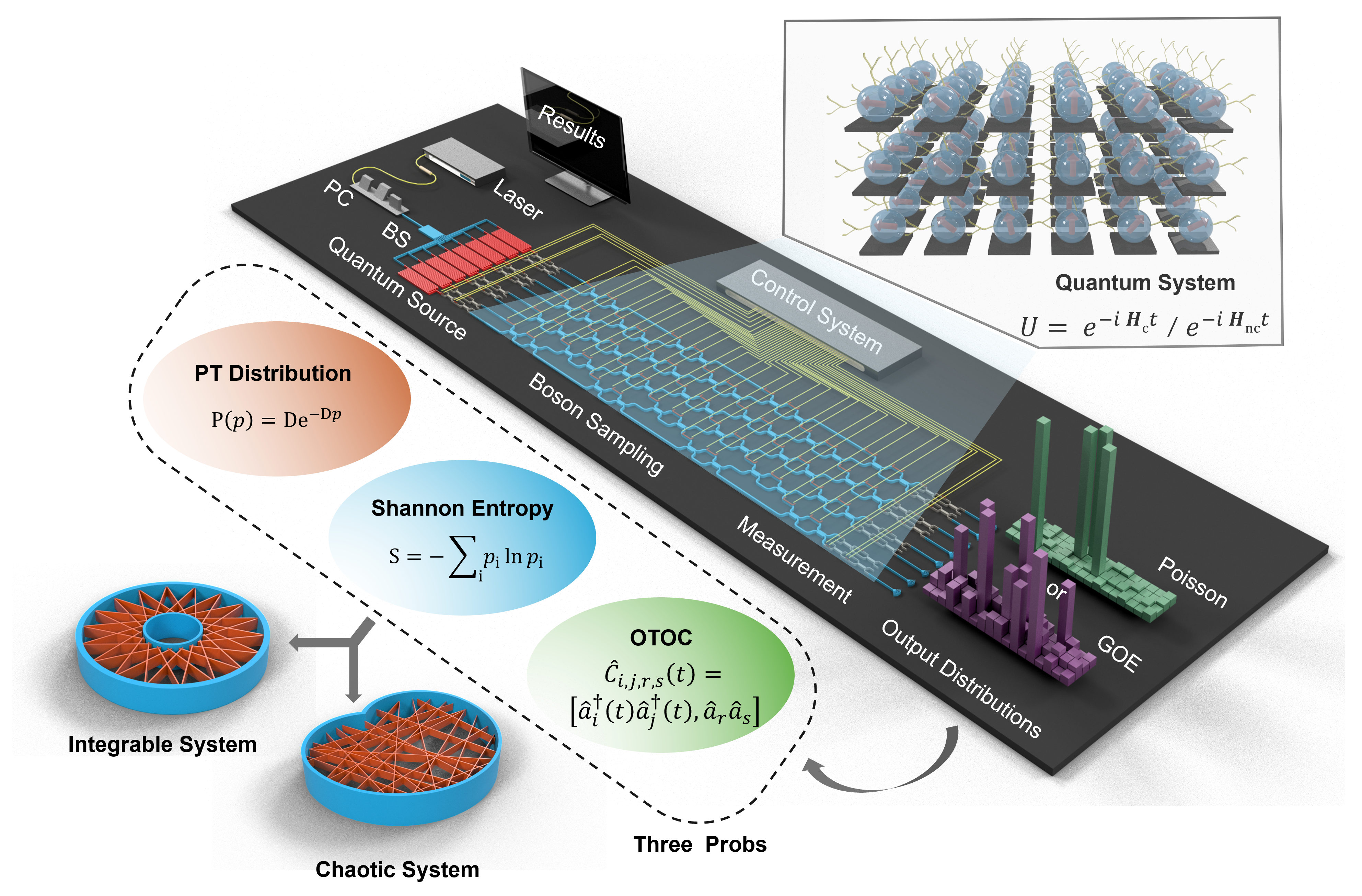}
  \caption{ \textbf{Workflow for distinguishing chaotic and integrable dynamics using on-chip boson sampling.}
  A quantum system undergoing unitary evolution $U=e^{-i H t}$ is mapped onto an integrated photonic platform, where $H_\text{c}$ and  $H_\text{nc}$ represent chaotic and integrable dynamics, respectively. The boson sampling experiment is implemented using a pump laser, polarization controller (PC), beam splitter (BS), and spiral waveguides for photon-pair generation, followed by a programmable optical network and correlation measurements. The measured output probability distributions are then used to extract three quantitative probes: distance from PT distribution (\eqref{eqn: closeness of empirical prob dist of probabilities to PT}), the Shannon entropy (\eqref{eqn: empirical average shannon entropy}), and OTOC-equivalent observables (\eqref{eqn: 4-point OTOC}). Together, these metrics provide a comprehensive framework for distinguishing chaotic and integrable quantum dynamics.}
  \label{fig:mainDiag}
\end{figure*}

Figure~\ref{fig:mainDiag} summarizes the overall workflow of this work. A quantum system undergoing unitary evolution is mapped onto a programmable integrated photonic platform, where Fock-state boson sampling is implemented through photon-pair generation, programmable optical transformations, and multiphoton correlation measurements. The time evolution of the resulting output probability distributions encodes information about the underlying dynamics and serves as experimentally accessible observables for probing quantum chaos. The measured statistics are subsequently analyzed using three complementary probes: distance from PT distribution, Shannon entropy, and OTOC-equivalent observables, providing a unified framework for distinguishing integrable and chaotic dynamics on programmable integrated photonic hardware.

\section{Theory}
\label{sec:setup}

\subsection{Boson Sampling}
\label{sub:boson_sampling}
To probe quantum dynamics using boson sampling, we consider an $N$-photon input state propagating through an $M$-mode interferometer. Programmable unitaries $U^{(l)}(t_k)\in \text{U}(M)$ generated by the Hamiltonian ensemble $H^{(l)}$ are implemented on an integrated photonic chip, where $l$ labels a given Hamiltonian realization and $t_k$ denotes the $k$-th sampled evolution time. The corresponding unitary evolution is given by $U^{(l)}(t)=e^{-iH^{(l)}t}$ in the optical-mode basis, where each mode corresponds to a basis state $|j\rangle=\hat a_j^\dagger\ket{0}=\ket{0_1,0_2,\dots,1_j,\dots,0_M}$ and $\hat a_j^\dagger$ denotes the creation operator of mode $j$. Bosonic interference between indistinguishable photons makes the measured output statistics highly sensitive to the underlying unitary dynamics and their associated random-matrix properties.

Based on this mapping, an $N$-photon input Fock state $\ket{\mathbf{n}^{\mathrm{in}}}=\ket{n_1^{\mathrm{in}},\dots,n_M^{\mathrm{in}}}$
evolves through the interferometer according to
$\hat a_i^\dagger(t_k)
  =
  \sum_{j=1}^{M}
  U_{ij}^{(l)}(t_k)
  \hat a_j^\dagger$. The probability of observing an output occupation pattern $\mathbf{n}^{\mathrm{out}}=(n_1^{\mathrm{out}},\dots,n_M^{\mathrm{out}})$ is then given by
\begin{equation}
\label{eq:OutputProb}
  p_{\mathbf{n}^{\rm out}}^{(l)}(t_k)
  =
  \frac{
  \left|
  \operatorname{Per}
  \left[
    U_{\mathbf{n}^{\rm out},\mathbf{n}^{\rm in}}^{(l)}(t_k)
    \right]
  \right|^2
  }{
  \prod_i n_i^{\rm out}! \prod_j n_j^{\rm in}!
  },
\end{equation}
where $U_{\mathbf{n}^{\rm out},\mathbf{n}^{\rm in}}^{(l)}(t_k)$ is the $N \times N$ submatrix constructed by repeating rows according to the output occupations $\mathbf n^{\rm out}$ and columns according to the input occupations $\mathbf n^{\rm in}$. For convenience, we arrange all the probabilities $p_{\mathbf n^{\rm out}}^{(l)}(t_k)$ for different output configurations $\mathbf n^{\rm out}$ into a probability vector $\vec p^{(l)}(t_k)$ for a given realization $l$ and time $t_k$. Due to the permanent ($\operatorname{Per}$) of $U_{\mathbf{n}^{\rm out},\mathbf{n}^{\rm in}}^{(l)}(t_k)$ appearing in~\eqref{eq:OutputProb}, the output probabilities become highly nontrivial functions of the underlying unitary evolution. We retain only two-click collision-free events and construct conditional probability distributions over collision-free output configurations. This approach avoids the need for photon-number-resolving detection while preserving the relevant statistical properties of the output distributions. As shown in Appendix~A,  this restriction preserves the relevant statistical properties of the conditional probability distributions used throughout this work, provided that the collision-free fraction $D/N_0$ remains sufficiently close to unity, where $D=\binom{M}{N}$
and $N_0=\binom{M+N-1}{N}$ denote the number of collision-free and total output configurations, respectively. This condition is satisfied in our experiment with
$D/N_0\simeq0.78$ (see Eq.~S2 and Eq.~S3).

\subsection{Random-Matrix Hamiltonians for Quantum Dynamics}
\label{sub:Chavda}
To generate integrable and chaotic dynamics, we employ a random-matrix Hamiltonian family~\cite{scaling_rm_PRA_1991,chavda2014transition},
\begin{equation}
  \label{eqn: chavda model}
  H_{\Lambda} = \frac{H_0 + \lambda V}{\sqrt{1+\lambda^2}},
\end{equation}
where $H_0$ is a diagonal matrix with independent Gaussian entries. The diagonal entries are $(H_0)_{ii} \sim \mathcal{N}(0,\,1)$, where $\mathcal{N}(\mu,\,\sigma^2)$ is the normal distribution with mean $\mu=0$ and variance $\sigma=1$, while $V$ is a $M \times M$ real symmetric matrix drawn from the GOE ensemble, with variances $2/M$ and $1/M$ for diagonal and off-diagonal elements, respectively. The perturbation strength $\lambda$ is commonly expressed through the dimensionless parameter $\Lambda=\lambda^2M/2\pi$, which characterizes the relative strength of the integrability-breaking perturbation with respect to the mean level spacing.
$H_0$ describes an integrable system with uncorrelated energy levels, whereas the GOE perturbation $V$ introduces level mixing and repulsion, driving the system toward chaotic behavior.

For $\Lambda\ll1$, level correlations between energy levels are weak and the system exhibits integrable (Poissonian) statistics, whereas for
$\Lambda\gg1$, strong spectral correlations lead to chaotic (GOE) behavior. Accordingly, varying $\Lambda$
drives a continuous crossover from Poissonian to GOE statistics. In this work, we focus on the two limiting cases, while intermediate regimes remain experimentally accessible.

We sample ensembles $\{H_0^{(l)}\}$ and $\{V^{(l)}\}$ and consider two representative values, $\Lambda=0.01$ and $\Lambda=1000$, defining Hamiltonian ensembles $\{H_{\mathrm{nc}}^{(l)}\}$ and $\{H_{\mathrm{c}}^{(l)}\}$ in the integrable  and chaotic  regimes, respectively. For each realization $l$, the Hamiltonians generate the evolution at five discrete time points $t_k\in[t_1,\dots,t_5]$, generating two ensembles of unitary matrices, $\{U_{\mathrm{nc}}^{(l)}(t_k)=e^{-iH_{\mathrm{nc}}^{(l)}t_k}\}$ and $\{U_{\mathrm{c}}^{(l)}(t_k)=e^{-iH_{\mathrm{c}}^{(l)}t_k}\}$ (see Appendix~B for detailed settings). These unitary matrices are then programmed onto the photonic chip, where boson sampling is performed using a fixed input state to obtain the corresponding output probability vectors $\{\mathbf p_{\mathrm{nc}}^{(l)}(t_k)\}$ and $\{\mathbf p_{\mathrm{c}}^{(l)}(t_k)\}$. For each evolution time, the probe quantities are computed from the realization-dependent output distributions and subsequently averaged over the corresponding ensembles. The three probes and the corresponding experimental results are defined and discussed in Section~\ref{sec:detection_model}. Although the implemented dynamics are generated through single-particle unitary evolution, bosonic indistinguishability produces nontrivial multiphoton interference at the output, allowing signatures of the underlying random-matrix dynamics to emerge in the measured probability distributions.
Our experimental implementation employs the random-matrix Hamiltonian family $H_\Lambda$ because of its theoretical simplicity and controllability. More generally, the proposed framework can be naturally extended to more general many-body spin Hamiltonians (see Appendix~C for further details).

\section{Experimental Implementation}
\label{sec:experiment}
\begin{figure*}
  \centering
  \includegraphics[width=0.95\linewidth]{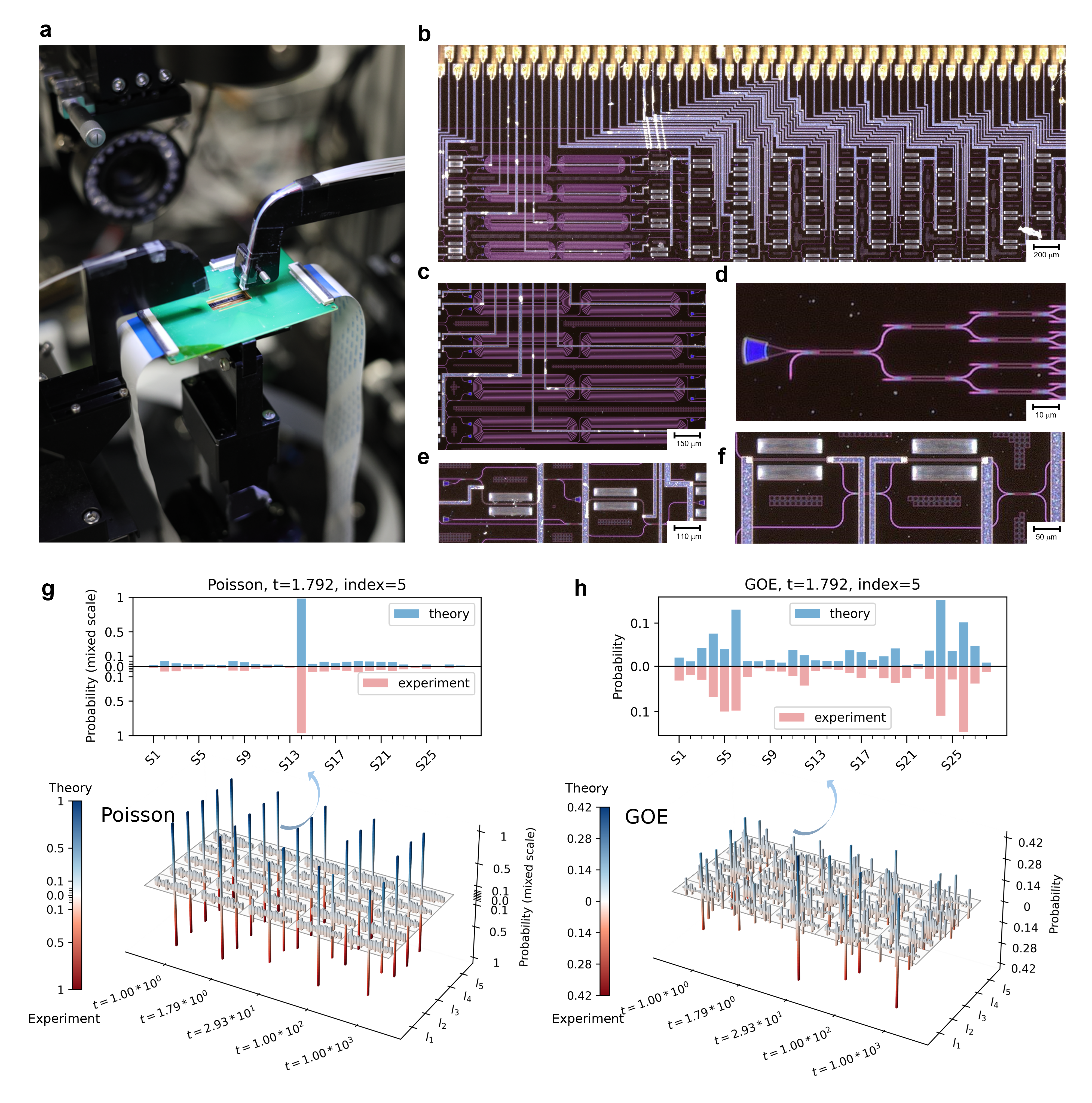}
  \caption{ \textbf{Experimental implementation of the on-chip boson sampling platform.}
  \textbf{a)} Schematic overview of the experimental setup integrating photon generation, manipulation, and measurement on a single photonic chip.
  \textbf{b)}  The 8-mode reconfigurable photonic circuit.
  \textbf{c)}  Spiral waveguide structures used for photon-pair generation via SFWM.
  \textbf{d)}  1-to-8 beam splitter network for distributing input light.
  \textbf{e)}  AMZI is used for pump filtering.
  \textbf{f)}  MZI structure.
  \textbf{g)}  Integrable dynamics of Hamiltonian $H_\text{nc}$. Top: Enlarged view of the output probability distribution $p_{\mathbf{n}^{\rm out}}^{(l)}(t_2)$ at $t^*=t_2=1.79$ for different outputs $\mathbf{n}^{\rm out}$, where blue and red bars denote the theoretical and experimental results, respectively. For visualization purposes, a mixed scale is used, with probabilities between 0 and 0.1 displayed on a logarithmic scale and probabilities between 0.1 and 1 displayed on a linear scale. Bottom: Experimental and theoretical output distributions for five randomly selected realizations at evolution times $t\in[1,1.79,29.29,100,1000]$.
  \textbf{h})  Chaotic dynamics of Hamiltonian $H_\text{c}$. Same as in \textbf{g)}, but for the GOE Hamiltonian ensemble.
  }
  \label{fig: exp}
\end{figure*}
In the experiment, we designed and fabricated an integrated photonic chip capable of integrating photon generation, manipulation, and measurement within a compact and reconfigurable architecture. The overall experimental setup is shown in Fig.~\ref{fig: exp}\textbf{a}, following a comprehensive characterization of the integrated photonic system (the detailed experimental procedure is described in Appendix~D.1).
Figure~\ref{fig: exp}\textbf{b} illustrates our experimental quantum photonic circuit. The 8-mode integrated chip, measuring 10 mm × 3 mm, provides a compact architecture with 100 independently adjustable phase shifters. Details of the calibration of these adjustable phases are discussed in Appendix~D.2. Each phase shifter is controlled externally via digital-to-analog converters (DACs), with connections made through two rows of metal pads bonded by gold wires on the upper and lower sides of the printed circuit board. This architecture is compatible with larger programmable photonic systems and therefore provides a pathway toward scaling to higher-dimensional implementations.

Photon pairs are generated through spontaneous four-wave mixing (SFWM) in eight spiral-waveguide quantum sources, as shown in Fig.~\ref{fig: exp}\textbf{c}.  To excite these sources simultaneously, the pump light is distributed into eight channels through a 1-to-8 beam splitter network (Fig.~\ref{fig: exp}\textbf{d}), forming an effective eight-input configuration. The photon sources exhibit high interference visibility, with detailed characterization and error analysis provided in Appendices~D.3 and~D.4, respectively. By tuning the pump power using MZIs placed before the quantum source, each source achieves a controllable photon generation rate of approximately $500~\mathrm{kHz}$.
Following generation, residual pump photons at wavelengths around 1545 nm and 1555 nm are filtered using the Asymmetric Mach–Zehnder interferometer (AMZI), as illustrated in Fig.~\ref{fig: exp}\textbf{e}. This filtering stage suppresses unwanted pump light that could otherwise degrade interference in the subsequent optical circuit. The filtered photons are then injected into a reconfigurable optical network capable of implementing arbitrary unitary transformations via the Clements decomposition protocol~\cite{clements2016optimal}. This approach decomposes any unitary matrix into a sequence of tunable MZIs (Fig.~\ref{fig: exp}\textbf{f}), achieving an average unitary fidelity of 0.93 (see Appendix~D.5 for details).  This circuit forms the computational core of quantum chaos experiments, with high flexibility in parameter selection.
Finally, photons exiting the circuit are coupled off-chip through a fiber grating array and detected using high-efficiency superconducting nanowire single-photon detectors. The total optical loss from chip to detector is approximately $10~\mathrm{dB}$, while the detector efficiency reaches $\sim 88\%$, with a dark count rate of about $50~\mathrm{s}^{-1}$. In data acquisition, only events where the number of detected photons matches the number of injected photons are retained for analysis (see Appendix~A for details).
Since our diagnostics rely on statistical properties of the output distribution rather than exact amplitudes, they are robust to moderate deviations from the ideal unitary, provided multiphoton indistinguishability and interference visibility remain high.

Boson sampling experimental results are shown in Figs~\ref{fig: exp}\textbf{g,h} for the Poissonian and GOE cases, respectively, corresponding to the integrable and chaotic ensembles introduced in Section~\ref{sub:Chavda}. For each evolution time point $t \in [1,1.79,29.29,100,1000]$, we show five randomly selected realizations, while detailed settings are provided in Appendix~B. The full output distribution consists of 28 two-photon configurations $\mathbf{n}^{\rm out}$, labeled as $s_1$ to $s_{28}$ (e.g., $s_1=\ket{1_1,1_2,0_3,\dots,0_8}$, $s_2=\ket{1_1,0_2,1_3,\dots,0_8}$, ..., $s_{28}=\ket{0_1,\dots,0_6,1_7,1_8}$). From the results, we observe that for all evolution times, the experimental measurements (red) are in good agreement with the theoretical predictions (blue), demonstrating the accuracy of the experimentally obtained output statistics. The sources of residual discrepancies include device imperfections and experimental noise contributions, detailed in Appendix~D.4. In addition, clear differences between the Poissonian and GOE cases can already be observed from the measured output distributions. In the Poissonian regime, only a few output configurations carry significant probability weight, while most remain close to zero, indicating a more localized distribution. In contrast, the GOE case exhibits probability spreading over more output configurations, reflecting stronger delocalization expected for chaotic dynamics. This qualitative distinction provides an intuitive picture for the quantitative diagnostic analysis presented in Section~\ref{sec:detection_model}.

\section{Results}
\label{sec:detection_model}
Using the experimentally measured boson sampling distributions defining the vector $\{\mathbf p^{(l)}(t)\}$ introduced above, we now analyze how signatures of quantum chaos emerge in the output statistics. Our objective is to distinguish integrable and chaotic dynamics through ensemble-averaged properties of the measured distributions. To this end, we employ three complementary probes: the proximity to PT distribution, the Shannon entropy, and OTOC-equivalent observables derived from output probabilities. Together, these quantities provide a unified characterization of the statistical, dynamical, and delocalization properties associated with quantum chaos.

\begin{figure*}[ht]
  \centering
  \includegraphics[width=0.9\linewidth]{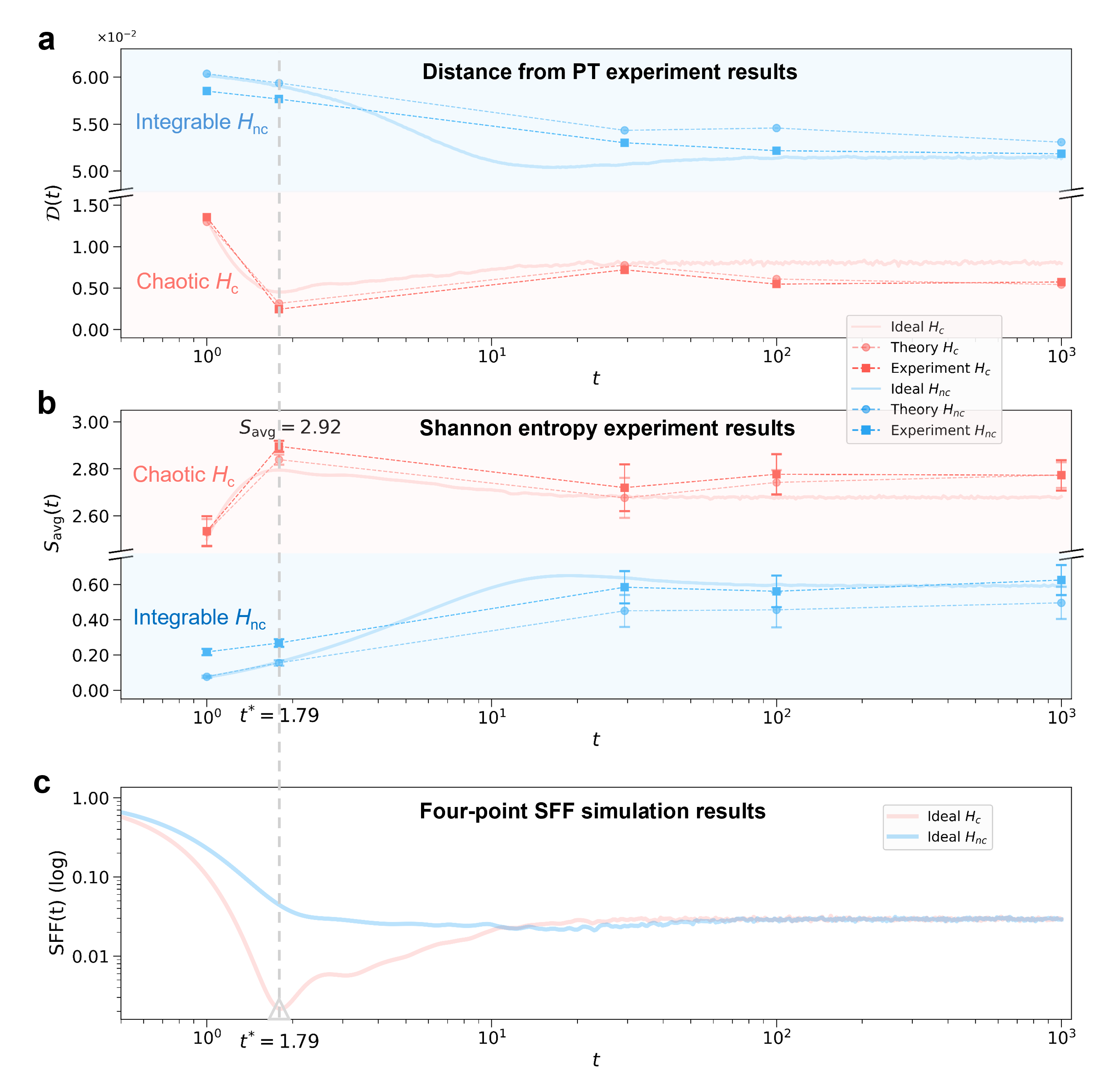}
  \caption{\textbf{Experimental demonstration of PT distributions and Shannon-entropy, and correspondence simulation results of $4$-point SFF.}
    \textbf{a)} The Wasserstein-1 distance $\mathcal{D}(t)$ for both chaotic ($H_\text{c}$, red region) and integrable ($H_\text{nc}$, blue region) dynamics. Experimental results
    (rectangular points) are obtained by performing boson sampling on the
    photonic chip (see Appendix~B for more details). The corresponding theoretical results (circle points) use the same unitary ensembles and sampling conditions. 
    The ideal curves (solid line) correspond to large-scale numerical simulations with $n_t=2000$ realizations and densely sampled evolution times. The close agreement between experiment and theory confirms the validity of the photonic implementation. The difference between the theoretical and ideal curves is primarily due to finite-sampling effects associated with the limited number of realizations. 
     \textbf{b)} The averaged Shannon entropy $S_{\rm avg}(t)$ for $H_\text{c}$ and $H_\text{nc}$. The symbols and colors follow those in \textbf{a}. Error bars in the Shannon entropy panel represent the standard error of the mean arising from the finite number of Hamiltonian realizations in \eqref{eqn: chavda model}.   
    \textbf{c)} Numerically simulated four-point SFF for the dynamics generated by $H_\text{c}$ and $H_\text{nc}$. The chaotic case exhibits the characteristic dip--ramp--plateau structure, with a minimum at $t^*=1.79$. We note that the minimum of the SFF is in direct correspondence with the dip in $\mathcal{D}(t)$ and the peak in $S_\text{avg}(t)$, as expected.
  }
  \label{fig: main_result PT SE}
\end{figure*}
\subsection{Distance from Porter--Thomas Distribution}
\label{sub:pt}
For each evolution time $t$, the boson sampling experiment produces a set of output probability distributions defining the vectors $\{\mathbf p^{(l)}(t)\}$ over different Hamiltonian realizations $l=1,\dots,n_t$ ( $n_t$ is the number of realizations at time $t$). To characterize their statistical properties, we construct the empirical distribution $f_t(p)$ of probability values by pooling all collision-free output probabilities across realizations and output configurations,
\begin{equation}
  f_t(p) = \frac{1}{N_{\mathrm{tot}}} \sum_{l=1}^{n_t}\sum_{i=1}^{D} \delta\big(p - p_i^{(l)}(t)\big),
  \label{eqn: probability dist of probabilities}
\end{equation}
where $i=1,\dots,D$ labels the allowed output configurations $\mathbf n^{\rm out}$ and $N_{\rm tot}=n_tD$ denotes the total number of sampled probabilities. The distribution $f_t(p)$ therefore represents the empirical histogram of output probability values.

A key finding of our work is the connection between the spectral structure of quantum chaos and experimentally accessible boson sampling statistics. For chaotic Hamiltonians, the unitary evolution $e^{-iHt}$ is expected to approach Haar-random behavior at intermediate times, corresponding to the minimum of the spectral form factor (SFF). We refer the reader to Appendix~E for a detailed explanation. Since boson sampling with Haar-random unitaries produces output probabilities following the Porter--Thomas distribution $P_{\mathrm{PT}}(p)=De^{-Dp}$ on average, the empirical probability distribution $f_t(p)$ is expected to approach PT in the large-$D$ limit when the SFF reaches a minimum (with finite-$D$ corrections discussed in Appendix~A). This motivates using the distance between the measured distribution and the Porter--Thomas distribution as an experimentally accessible probe of chaos. To quantify this proximity, we compute the Wasserstein-1 distance~\cite{panaretos2019statistical}  between the empirical distribution $f_t(p)$ and the Porter--Thomas distribution $P_{\mathrm{PT}}(p)$
\begin{equation}
  \label{eqn: closeness of empirical prob dist of probabilities to PT}
  \mathcal{D}(t) = W_1\big[f_t(p), P_{\mathrm{PT}}(p)\big] \ ,
\end{equation}
where $W_1$ is calculated from the corresponding cumulative distribution functions. The Wasserstein distance is particularly suitable because it directly measures the global discrepancy between two probability distributions and remains robust for finite-sample experimental data. Hence, for chaotic Hamiltonians, we expect that $\mathcal{D}(t)$ should exhibit a characteristic dip as a function of time, corresponding to the minimum of the SFF, followed by a plateau at longer times. In contrast, for integrable dynamics no such feature is expected.

The experimental and theoretical results are shown in Fig.~\ref{fig: main_result PT SE}\textbf{a}, together with ideal numerical curves obtained from larger ensembles and more densely sampled evolution times for comparison. For chaotic dynamics, $\mathcal{D}(t)$ exhibits a pronounced dip and approaches values close to zero around $t^*=1.79$, which coincides with the minimum of the SFF shown in Fig.~\ref{fig: main_result PT SE}\textbf{c}. This agreement confirms the theoretical expectation that chaotic dynamics become increasingly similar to Haar-random behavior near the SFF dip region. Furthermore, we see a ramp and a plateau after the dip, corresponding to the SFF behavior as depicted in Fig.~\ref{fig: main_result PT SE}\textbf{c}. In contrast, integrable dynamics consistently remain farther from the PT distribution throughout the evolution. Therefore, the proximity of the output statistics to PT behavior provides a direct experimental signature for distinguishing chaotic and integrable dynamics. Given the finite system size and limited number of time points in our experiment, we interpret the observed behavior of $\mathcal{D}(t)$ as a qualitative indicator rather than a precise reconstruction of the spectral form factor.

\subsection{Shannon Entropy}
\label{sub:se}

The second complementary probe is the Shannon entropy of the output probability distribution. For each realization $l$ and time $t$, the entropy of the corresponding probability vector $\mathbf p^{(l)}(t)$ is defined as
\begin{equation}
  S^{(l)}(t) = - \sum_{i=1}^D p_i^{(l)}(t)\,\ln p_i^{(l)}(t),
  \label{eqn: D shannon entropy}
\end{equation}
where the summation runs over all collision-free output configurations. To characterize the ensemble behavior, we compute the average entropy over all $n_{t_k}$ realizations at each evolution time $t_k$,
\begin{equation}
  S_{\text{avg}}(t_k) = -\frac{1}{n_{t_k}} \sum_{l=1}^{n_{t_k}} \sum_{i=1}^D p_i^{(l)}(t_k)\,\ln p_i^{(l)}(t_k),
  \label{eqn: empirical average shannon entropy}
\end{equation}
which quantifies how uniformly the probability weight is distributed across the accessible output space. Lower entropy indicates that the probability distribution is concentrated on a small number of configurations, whereas larger entropy corresponds to stronger delocalization over the accessible Hilbert space.

The central observation underlying this probe is that chaotic dynamics lead to stronger spreading of probability weight across the output space than integrable dynamics. Since chaotic evolution approaches Haar-random behavior near the dip region of the spectral form factor, the corresponding boson sampling distributions are expected to become maximally delocalized, resulting in a peak in the Shannon entropy. Detailed derivations are provided in Appendix~E. For Haar-random unitaries, the expected entropy can be evaluated analytically as
\begin{equation}
  \langle S \rangle_{\mathrm{Haar}} \;=\; -1 + \sum_{i=1}^{D}\frac{1}{i},
  \label{eq:haar-entropy}
\end{equation}
and this gives predicted maximum entropy
\begin{equation}
  \max_{t} S_{\mathrm{avg}}(t) \;\approx\; -1 + \sum_{i=1}^{D}\frac{1}{i}
  \;\approx\; 2.92\qquad (D=28) \ .
  \label{eq:Smax}
\end{equation}

The experimental and theoretical results are shown in Fig.~\ref{fig: main_result PT SE}\textbf{b}. For chaotic dynamics, $S_{\mathrm{avg}}(t)$ exhibits a pronounced peak around $t^*=1.79$, where the experimentally observed value approaches the predicted Haar limit of $2.92$. This behavior is consistent with the expectation that chaotic evolution produces increasingly delocalized output distributions near the SFF dip region. In contrast, the entropy of integrable dynamics remains consistently lower throughout the evolution, indicating that the probability weight remains concentrated on a smaller subset of output configurations.
Residual discrepancies between the experimental results and the predicted Haar limit mainly arise from finite-size effects and experimental limitations. Numerical simulations in Appendix~F show that these finite-size effects decrease with increasing system size, leading to improved agreement with the Haar prediction and a clearer separation between chaotic and integrable dynamics. Additional deviations originate from experimental imperfections, including finite sampling statistics and hardware imperfections, which are analyzed in Appendix~E.4.

Therefore, the Shannon entropy of the boson sampling output distribution provides a second experimental signature for distinguishing chaotic and integrable dynamics. Physically, this probe quantifies the degree of probability spreading in the accessible Hilbert space and provides a direct measure of the delocalization induced by chaotic dynamics.

\subsection{OTOC-equivalent Observables}
\label{sub:otoc}

The third complementary probe of quantum chaos is constructed from OTOC-equivalent observables derived from boson sampling statistics. Instead of directly measuring conventional OTOCs, we exploit a mapping between boson sampling output probabilities and four-point correlators to obtain experimentally accessible signatures of information spreading and delocalization. This approach enables the dynamics of quantum chaos to be probed directly from measured output probabilities on a programmable photonic platform. To establish this connection, we use the mapping derived in our work~\cite{bastidas2025equilibration}, where the output probabilities of boson sampling with a two-photon Fock input are shown to be equivalent to four-point OTOCs of single-particle creation and annihilation operators.
In particular, for a two-photon input state $\hat{a}_i^\dagger \hat{a}_j^\dagger |0\rangle$, the probability of detecting photons in output modes $r$ and $s$ can be expressed as
\begin{equation}
  \label{eqn: 4-point OTOC}
  C^{(4)}_{i,j,r,s}(t) = \langle 0 | \hat{C}^\dagger_{i,j,r,s}(t)\, \hat{C}_{i,j,r,s}(t) |0\rangle,
\end{equation}
where $\hat{C}_{i,j,r,s}(t) = [\hat{a}_i^\dagger(t)\hat{a}_j^\dagger(t), \hat{a}_r \hat{a}_s]$
and $\hat{a}_i^\dagger(t) = \sum_j U^{(l)}_{ij}(t)\hat{a}_j^\dagger$ is the time-evolved creation operator in the Heisenberg picture. These configuration-resolved output probabilities therefore act as OTOC-equivalent observables that probe the delocalization of single-particle excitations across optical modes through bosonic interference. Since the implemented dynamics originate from single-particle unitary evolution, these observables characterize bosonic delocalization rather than conventional many-body scrambling in interacting systems (we retain the standard OTOC terminology for convenience).

\begin{figure*}[ht]
  \centering
  \includegraphics[width=0.99\textwidth]{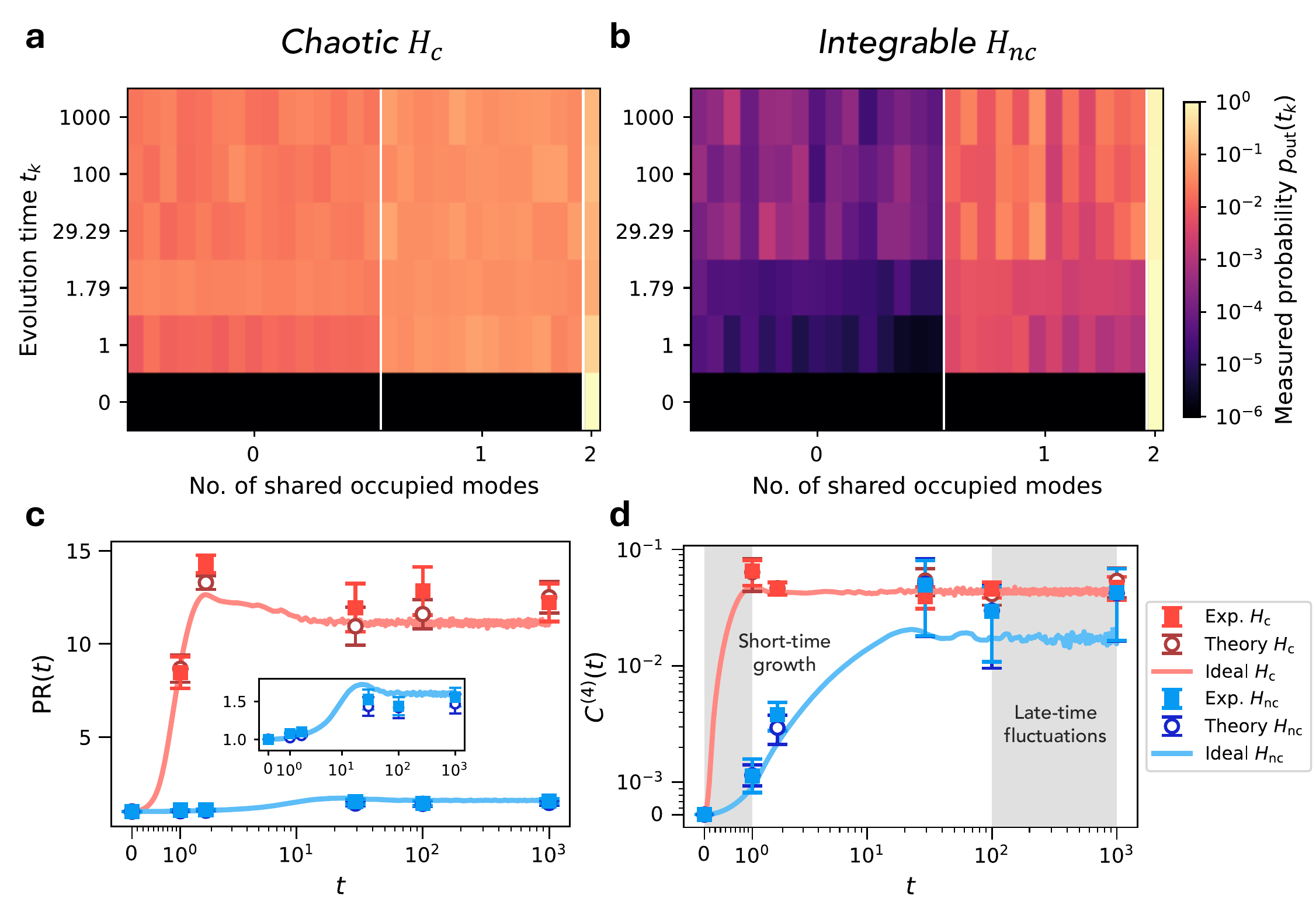}
  \caption{\textbf{Experimental demonstration of OTOC-equivalent observables as a probe of quantum chaos.}
    \textbf{a,b)} Ensemble-averaged probabilities for all output configurations measured under \textbf{a)} chaotic $H_\text{c}$ and \textbf{b)} integrable $H_\text{nc}$ dynamics, grouped by the number of occupied modes shared with the initial configuration. The sector with two shared occupied modes corresponds to the input configuration. \textbf{c)} Participation ratio $\text{PR}(t)$ for chaotic $H_\text{c}$ (red) and integrable $H_\text{nc}$ (blue) dynamics. The inset shows a magnified view of the integrable case. \textbf{d)} Representative four-point OTOC-equivalent observable for the output configuration $\ket{0_1,0_2,1_3,0_4,0_5,1_6,0_7,0_8}$ under chaotic $H_\text{c}$ (red) and integrable $H_\text{nc}$ (blue) dynamics. Shaded regions indicate the short-time growth and late-time fluctuation regimes. In \textbf{c)} and \textbf{d)}, solid squares denote experimental data, open circles denote theoretical results obtained from the same unitary ensembles, and solid curves denote ideal averages over 2000 realizations. Each data point represents the mean over all corresponding $n_{t_k}$ realizations, with error bars denoting the standard error due to the finite number of instances. For all panels, $t_k=0$ is not experimentally measured since the only possible output is the input configuration.}
  \label{fig: experimental_OTOC}
\end{figure*}


Figures~\ref{fig: experimental_OTOC}\textbf{a,b} show the ensemble-averaged probabilities for all measured output configurations under chaotic and integrable dynamics. The output configurations ($s_1,\dots,s_{28}$) are grouped according to the number of occupied modes shared with the initial state $\ket{0_1,0_2,1_3,1_4,0_5,0_6,0_7,0_8}$. Since the random-matrix Hamiltonian ensemble lacks an underlying notion of spatial locality and instead exhibits effectively all-to-all mode couplings, the output statistics are more naturally organized according to overlap with the input configuration rather than by output-mode distance. This overlap structure provides a direct measure of memory retention and information spreading: configurations sharing more occupied modes with the input state retain stronger memory of the initial configuration and therefore generally exhibit larger probabilities, whereas configurations with smaller overlap correspond to stronger scrambling and lower probabilities. This hierarchy arises naturally from perturbative considerations, since configurations with larger overlap require fewer off-diagonal transitions from the initial state and therefore acquire larger probabilities at early times (see Appendix~G for further details).

To characterize the global spreading behavior beyond individual output configurations, we consider the participation ratio (PR)~\cite{participation_ratio_OTOC_lea,Victor_quantum_metamorphism}
\begin{equation}
  \label{eqn: participation_ratio}
  \mathrm{PR}^{(l)}(t) = \frac{1}{\sum_{i=1}^D \big[p_i^{(l)}(t)\big]^2},
\end{equation}
where the summation runs over all collision-free output configurations. The PR measures the effective number of configurations contributing significantly to the output distribution. For example, a fully localized distribution gives $\mathrm{PR}=1$, whereas a highly delocalized distribution over $D$ configurations gives $\mathrm{PR}\sim D$.
Fig.~\ref{fig: experimental_OTOC}\textbf{c} shows the ensemble-averaged PR as a function of time for chaotic and integrable dynamics. Clear differences can be observed between the two regimes. In the chaotic case, the participation ratio rapidly increases and reaches values around $12$, indicating that probability weight spreads across a large fraction of the accessible configuration space. In contrast, the integrable case remains close to unity, with values around $1.6$, reflecting a much stronger retention of probability weight within a limited subset of configurations. Experimental measurements show good agreement with theoretical predictions across the entire evolution. Thus, PR provides an effective global measure of Hilbert-space delocalization and serves as an additional experimental signature for distinguishing chaotic and integrable dynamics.

For a representative configuration, Fig.~\ref{fig: experimental_OTOC}\textbf{d} shows the temporal evolution of the ensemble-averaged probability (OTOC-equivalent observable) of observing the output state $\ket{0_1,0_2,1_3,0_4,0_5,1_6,0_7,0_8}$ from the input state $\ket{0_1,0_2,1_3,1_4,0_5,0_6,0_7,0_8}$ for both chaotic and integrable dynamics. The representative dynamics exhibit the characteristic regimes commonly associated with OTOC dynamics: an initial growth regime followed by saturation and late-time fluctuations around a steady-state value. The close agreement between experiment and theory shows that the measured boson sampling probabilities accurately reproduce the predicted OTOC-equivalent dynamics. In Appendix~G, we further analyze these observables theoretically and numerically, showing that the short-time scaling follows overlap-dependent power laws that are largely independent of the underlying dynamics, whereas the late-time fluctuations provide additional signatures distinguishing chaotic and integrable behavior.

These observations are consistent with the behavior identified in Sections~\ref{sub:pt} and~\ref{sub:se}. In particular, the growth and saturation of the PR coincide with the minimum of the PT distance $\mathcal{D}(t)$ and the peak of the Shannon entropy. Together, these three probes provide complementary perspectives on the same underlying dynamics: the Porter--Thomas distance characterizes the emergence of random-matrix behavior, the Shannon entropy quantifies probability delocalization, and the OTOC-equivalent observables capture information scrambling. Therefore, they provide a unified picture in which chaotic dynamics lead to rapid spreading of probability weight, reduced memory of the initial configuration, and enhanced scrambling across the accessible Hilbert space.

\section{Discussion and Conclusion}
From a theoretical perspective, the consistency among the three proposed probes suggests that different statistical quantities extracted from boson sampling outputs capture common signatures of the underlying chaotic dynamics. The distance from the PT distribution characterizes the emergence of random-matrix behavior, the Shannon entropy quantifies Hilbert-space delocalization, and the OTOC-equivalent observables probe information scrambling. Although these probes capture different physical aspects of the dynamics, they exhibit correlated temporal behavior and consistently identify the same underlying dynamical transition.
Our results also establish an operational connection between boson sampling statistics and conventional measures of quantum chaos. If the underlying Hamiltonian is already believed to be chaotic, then boson sampling data can be used to estimate where the dynamics come closest to Haar-like behavior. In the language of spectral statistics, this gives experimental access to the time where the spectral form factor reaches its minimum value, without having to reconstruct the full spectrum or to use time-reversal protocols. Hence, we show that multiphoton interference provides an operational way to locate the onset of strongest scrambling in a system whose microscopic dynamics may otherwise be difficult to characterize directly. Taken together, these results establish a direct correspondence between experimentally accessible boson sampling statistics and established probes of quantum chaos, providing a unified framework that connects spectral properties, information scrambling, and multiphoton interference.

From an experimental perspective, the programmable integrated photonic chip provides several advantages for probing quantum chaos. The reconfigurable architecture enables the implementation of large ensembles of unitary dynamics and access to time-dependent boson sampling statistics, making it possible to experimentally probe spectral properties, Hilbert-space delocalization, and information scrambling within a unified framework. Since boson sampling intrinsically relies on multiphoton interference and computationally complex output statistics, the proposed framework naturally inherits key advantages of photonic quantum information processing for accessing complex dynamical behavior.
Finite system size and hardware imperfections, including loss and device nonidealities, primarily affect the sharpness of the asymptotic predictions while preserving the qualitative distinction between chaotic and integrable dynamics, consistent with both theoretical and experimental results.
Numerical scaling results further indicate that these signatures become increasingly pronounced with increasing system size (see Appendix~F), where larger systems are expected to provide stronger separation between ergodic and non-ergodic dynamics and enable more direct studies of information scrambling in higher-dimensional Hilbert spaces. These observations highlight the potential of integrated photonic boson sampling platforms for studying increasingly complex quantum dynamics.

Looking forward, although the present work focuses on a random-matrix Hamiltonian framework, the proposed approach is not restricted to this specific model. As discussed in Appendix~C, the same framework can be naturally extended to more general many-body spin systems through the Holstein--Primakoff mapping in the dilute regime. The current implementation therefore serves as a proof-of-principle demonstration of a broader methodology for probing quantum chaos through boson sampling statistics.
Beyond random-matrix dynamics, the proposed framework may enable future studies of a wider range of dynamical phenomena on programmable photonic platforms, including localization transitions, nonergodic behavior, and more general scrambling dynamics~\cite{kulkarni_photonic_PRA_2025}. 
Since boson sampling has found applications in areas ranging from graph-theoretic problems to molecular vibronic spectra, our results further raise the question of whether the computational hardness arising in such problems is related to the same mechanisms of delocalization and information spreading that underlie quantum chaos. The present work provides a concrete experimental framework for systematically investigating these potential connections in future studies.
More broadly, the present work highlights the potential of connecting multiphoton interference with signatures of complex quantum dynamics, providing new opportunities for studying nonequilibrium quantum physics and information dynamics in larger quantum systems. 

In conclusion, we proposed and experimentally demonstrated a boson-sampling-based framework for probing quantum chaos on a programmable integrated photonic platform. Theoretically, we establish an operational connection between boson sampling statistics and conventional signatures of quantum chaos, enabling experimentally accessible probes of spectral properties, Hilbert-space delocalization, and information scrambling directly from measured output distributions. Experimentally, we realize these ideas on an eight-mode silicon photonic processor with two indistinguishable photons, combining arbitrary unitary programming and multiphoton correlation measurements within a scalable integrated architecture.
Using three complementary probes extracted from boson sampling statistics, we show that chaotic and integrable dynamics exhibit clearly distinguishable signatures. The proximity to Porter--Thomas statistics reveals the emergence of Haar-like behavior and characteristic features of the spectral form factor; the Shannon entropy captures enhanced probability spreading and Hilbert-space delocalization; and the OTOC-equivalent observables characterize stronger information scrambling and reduced memory retention in chaotic systems. Experimental results show good agreement with theoretical predictions and consistently distinguish chaotic and integrable dynamics across all measured observables.
Our work establishes boson sampling as a practical tool for studying quantum chaos and demonstrates the potential of integrated photonic platforms for investigating complex quantum dynamics. More broadly, the proposed framework provides new opportunities for exploring nonequilibrium quantum physics, information dynamics, and related phenomena in larger programmable quantum systems.

\newpage

\section*{Acknowledgements}
L.C.K. acknowledges support from the National Research Foundation, Singapore, and the Ministry of Education, Singapore. K.H.L acknowledges the Plan France 2030 through the project NISQ2LSQ (Grant ANR-22-PETQ-0006), the project OQuLus (Grant ANR-22-PETQ-0013), and also the CNRS@CREATE internal grant NGAP (NRF2023-ITC004-001). N.T.W.K. acknowledges support from the CQT NQSS PhD scholarship. K.C. gratefully acknowledges the Milner Foundation for their support. K.H.L would like to thank Dariel Mok Wai-Keong for helpful discussions.

\section*{Author contributions}
All authors contributed to the conception of the work. Y.C.Z. designed the chip, built the experimental setup, and performed the experiments. K.H.L., K.C., and N.T.W.K. equally contributed to the theoretical derivations, calculations, and simulations. H.Z. and L.X.W. assisted with the experiments, and V.M.B. assisted with the theoretical analysis. S.C., A.Q.L., Y.O., and L.C.K. supervised and coordinated all the work. Y.C.Z., K.H.L., K.C., and N.T.W.K. wrote the manuscript with contributions from all co-authors.

\section*{Data availability}
The data supporting the findings of this study are available from the corresponding author on reasonable request.

\section*{Competing interests}
The authors declare no competing interests.

\section*{Additional information}
The Appendix is available for this paper.

\newpage

\renewcommand{\theHsection}{appendix.\thesection}
\renewcommand{\theHsubsection}{appendix.\thesubsection}
\renewcommand{\theHfigure}{appendix.\thefigure}
\renewcommand{\theHequation}{appendix.\theequation}

\begin{appendices}

  \renewcommand{\thefigure}{S\arabic{figure}}
  \renewcommand{\thetable}{S\arabic{table}}
  \renewcommand{\theequation}{S\arabic{equation}}

  \setcounter{figure}{0}
  \setcounter{table}{0}
  \setcounter{equation}{0}

\section{Conditional Probabilities and Photon-Number Detection}
  \label{app: pnr detectors and conditional probabilities}
  In this appendix, we clarify the probability distributions considered throughout the main text and discuss their connection to the experimental detection scheme.

  \subsection{Experimental Detection and Post-Selection}

  In an ideal boson sampling experiment with photon-number-resolving detectors, one would have access to the full set of output configurations, including events where multiple photons occupy the same mode (collisions). For a system with $M$ modes and $N$ photons, the total number of possible output configurations is
  \begin{equation}
    N_0 = \binom{M + N - 1}{N} \ .
  \end{equation}
  In the experiment, we consider $M=8$ and $N=2$ , giving
  \begin{equation}\label{eq:s2}
    N_0 = \binom{8+2-1}{2} = 36 \ .
  \end{equation}
  However, due to the use of non-number-resolving detectors, we restrict attention to events in which the number of detected photons matches the number of injected photons, and in which no two photons occupy the same output mode. This corresponds to selecting the subset of collision-free outcomes, whose number is
  \begin{equation}\label{eq:s3}
    D = \binom{M}{N} = \binom{8}{2} = 28 \ .
  \end{equation}

  \subsection{Conditional Probability Distribution}

  Let $p_j$ denote the ideal probability of observing output configuration $j$, including both collision-free and collision events. Since only collision-free outcomes are retained, the conditional probability distribution is defined by
  \begin{equation}
    p^{\mathrm{cond}}_j = \frac{p_j}{\sum_{k \in \mathcal{C}} p_k}, \quad j \in \mathcal{C},
  \end{equation}
  where $\mathcal{C}$ denotes the set of collision-free configurations.
  This normalization ensures that the probabilities sum to one over the experimentally accessible subset:
  \begin{equation}
    \sum_{j \in \mathcal{C}} p^{\mathrm{cond}}_j = 1 \ .
  \end{equation}

  \subsection{Relation to the Main Text}

  The probability vectors $\mathbf{p}^{(l)}(t)$ should be interpreted as conditional distributions over collision-free outcomes in the main text,
  \begin{equation}
    \mathbf{p}^{(l)}(t) = \left( p^{(l)}_1(t), \ldots, p^{(l)}_D(t) \right) \ .
  \end{equation}
  All quantities defined in Section 4, including the empirical distribution $f_t(p)$, the Wasserstein distance $\mathcal{D}(t)$, the Shannon entropy $S_{\mathrm{avg}}(t)$, and the participation ratio $\mathrm{PR}(t)$, are computed using these conditional probability distributions.

  The following derivation shows that the probability distribution of the conditional probabilities follows the Porter--Thomas distribution when boson sampling is performed with a Haar-random unitary $U$. Specifically,
  $\text{probability}(p_j^\text{cond} \in [p, p+dp)) \approx D e^{-Dp} dp$ when $(N_0 - D)/N_0 \approx 0$.
  Since $U$ is a Haar random unitary, the probability distribution
  of $p_j$ follows the Porter--Thomas distribution. The probability distribution of $Y = 1 - \sum_{x  \in \mathcal{C}} p_x$ is given by
  \begin{equation}
    \text{Probability}(Y \in [y, y+ dy)) = \frac{N_0^C (1-y)^{C-1} e^{-N_0(1-y)}}{(C-1)!},
  \end{equation}
  where $C = N_0 - D$, using the fact that the sum of exponential random variables follows a Gamma distribution. $p_j^\text{cond}$ is essentially a ratio
  distribution between the Porter--Thomas distribution in $p_j$ and the Gamma
  distribution of $Y$. Thus, denoting $f(p)$ the probability distribution
  of $p_j^\text{cond}$, we have the following equation using the methods in~\cite{curtiss1941on}
  \begin{subequations}
    \label{eqn: porter thomas conditional probability approximation}
    \begin{align}
      f(p) & = \int_0^1 dy \,\, y N_0 e^{-N_0 p y} \frac{N_0^C (1-y)^{C-1} e^{-N_0(1-y)}}{(C-1)!} \\
           & \approx D e^{-Dp},
    \end{align}
  \end{subequations}
  where the approximation to a Porter--Thomas distribution in the last line is better for smaller values of $(N_0 - D)/N_0$.

\begin{figure}
    \centering
    \includegraphics[width=1\textwidth, trim={0cm 0cm 0cm 0cm},clip]{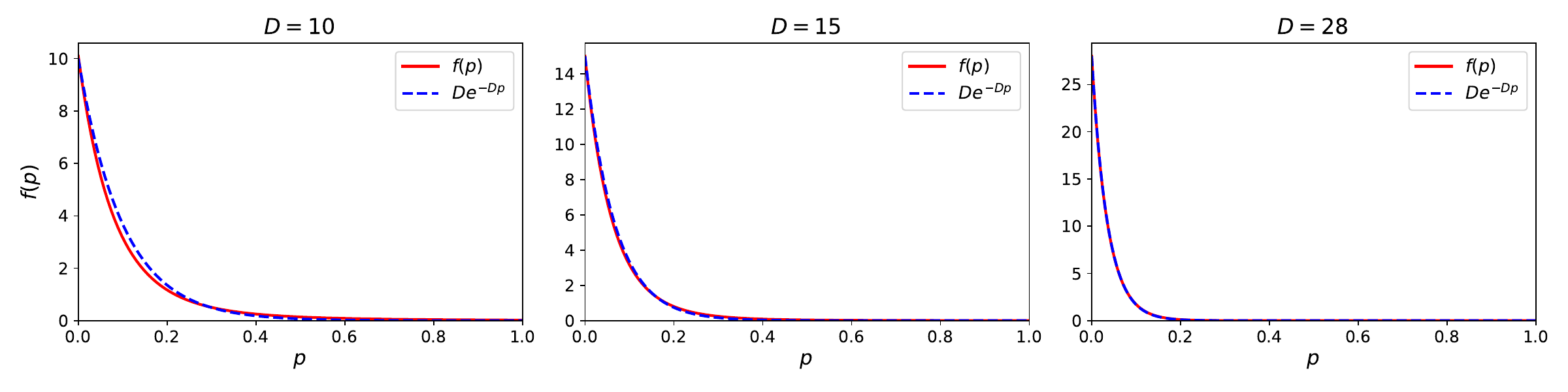}
    \caption[]{\textbf{$f(p)$ plots for various values of $D$.}
      For $N_0 = 36$ and $D = 28$,
      $p_j^\text{cond}$ follows the Porter--Thomas distribution perfectly, which justifies the use of the PT distribution as a substitute for the actual distribution of $p_j^\text{cond}$ in our analysis. We
      also show that for smaller values of $D$, this approximation (see \eqref{eqn: porter thomas conditional probability approximation}) of
      $p_j^\text{cond}$ by the PT distribution becomes less accurate.}
    \label{fig: conditional probability PT proof}
  \end{figure}

  The case $N_0=36$ with various values of $D$ is shown in Fig.~\ref{fig: conditional probability PT proof}.  Here, $N_0=36$ corresponds to the full Hilbert-space dimension of a boson-sampling experiment with two input photons and photon-number-resolving detectors, whereas $D=28$ corresponds to the experimental setting considered in this work, where events with a single detector click are discarded. These discarded events correspond to cases in which both photons exit through the same mode.
  As can be seen, for small values of $D$, the approximation is poor. By contrast, for $D = 28$, the approximation is essentially perfect, thereby justifying the use of the conditional probabilities in the Shannon Entropy formula. The results further indicate that this approximation becomes invalid for sufficiently small photonic systems. For example, in a three-mode photonic chip, the number of collision events equals the number of collision-free events, making the Porter--Thomas approximation for $p_j^\text{cond}$ inappropriate when collision events are discarded.

  \subsection{Remarks}

  The use of conditional probabilities reflects a standard experimental constraint in photonic boson sampling. While this restriction modifies the full output distribution, the statistical diagnostics considered in this work remain well-defined and robust, as they depend only on the relative distribution of probability weight across the accessible configurations.
A more complete treatment incorporating collision events would require photon-number-resolving detectors. Nevertheless, the inclusion of such events is not expected to qualitatively alter the distinction between chaotic and integrable dynamics observed in this work.

  \section{Construction of Unitary Ensembles for the Integrable--Chaotic Transition}
  \label{app: construction of unitaries}

  In the experiments, sixteen matrices were sampled for $H_0$ and $V$ and the same samples were used to construct $H_\text{nc} = H_{\Lambda = 0.01}$ and $H_\text{c} =
    H_{\Lambda = 1000}$. The chosen values of $\Lambda$ ensure that $H_\text{c}$ and $H_\text{nc}$ generate chaotic and integrable dynamics
  respectively~\cite{chavda2014transition}. The values $t_k$ selected are
  $[t_1,t_2,t_3,t_4,t_5] = [1,1.79,29.29,100,1000]$, and are chosen so as to
  see the behavior of $e^{-iH_\Lambda t}$ across a wide range of time values. Then, the matrices are sampled to construct the unitaries $e^{-i H_\text{c} t_k}$ and $e^{-i H_\text{nc} t_k}$ for $t_k = t_1, t_3, t_4, t_5$. Specially, for $t_k = t_2$, 
   seventy-five matrices are sampled for $H_\text{c}$ and sixteen matrices are sampled for $H_\text{nc}$.  We sampled more matrices for $t = t_2$ for $H_{c}$ as it is a theoretically important point, being the point at which the $4$-point SFF of $H_{c}$ reaches a minimum.  The importance of $t = t_2$ for $H_\text{c}$ is explained in Section~2.2.

 After constructing the realizations of the unitary matrices $e^{-i H_\text{c} t_k}$ and $e^{-i H_\text{nc} t_k}$ for the selected values of $t_k$, these unitaries are implemented on our photonic chip using the decomposition procedure described in Appendix~\ref{app: chip decomposition}. Boson-sampling measurements are subsequently performed according to the experimental protocol outlined in Appendix~\ref{app: experimental setup}.

\section{Generalization to Spin Hamiltonians}
  \label{app: spin Hamiltonian}

The proposed framework extends beyond the random-matrix model $H_\Lambda$ considered in this work and applies to a broad class of many-body spin Hamiltonians. Consider the following spin Hamiltonian:
  \begin{equation}
    H_s = \sum_{i,j=1}^K S_i H_{ij} S_j^\dagger,
  \end{equation}
  where $H_{ij}$ are matrix elements of a Hermitian matrix, $S_i$ is the
  spin-$1/2$ lowering operator on the $i$th spin, and $K$ is the total number of spins. The Hamiltonian $H_s$ commutes with the total magnetization operator
\begin{equation}
M = \sum_{i=1}^K S_i^z,
\end{equation}
where $S_i^z$ represents the projection of the $i-$th spin onto the $z$ axis. 
  The Holstein--Primakoff transformation,
  \begin{subequations}
    \begin{align}
      S_i   & = a_i^\dagger \sqrt{1 - a_i^\dagger a_i} \\
      S_i^z & = \frac{1}{2} - a_i^\dagger a_i,
    \end{align}
  \end{subequations}
  converts $H_s$ into a bosonic Hamiltonian $H_b$,
  \begin{equation}
    \label{eqn: after HP transform spin-1/2}
    H_b = \sum_{i,j=1}^K H_{ij} a_i^\dagger (1-a_i^\dagger a_i) a_j.
  \end{equation}
The Holstein--Primakoff transformation maps the dynamics generated by $e^{-iH_st}$ in the $m$-magnetization subspace to the dynamics generated by $e^{-iH_bt}$ in the subspace containing
\begin{equation}
N = K/2 - m
\end{equation}
photons, where $m$ is the eigenvalue of the total magnetization operator $M$.
        
For boson-sampling experiments satisfying $K \gg N^2$, the probability of observing two or more photons occupying the same mode becomes vanishingly small for typical unitaries as $K$ increases~\cite{aaronson2011computational,spagnolo2013general,peropadre2017equivalence}. Under this condition, 
  $a_i^\dagger \sqrt{1- a_i^\dagger a_i}\approx a_i^\dagger$,  and the Holstein--Primakoff mapping reduces to,
\begin{equation}
H_s
\overset{\text{Holstein--Primakoff}}{\longrightarrow}
H_b
\approx
\sum_{i,j} H_{ij} a_i^\dagger a_j.
\end{equation}
Consequently, boson sampling with $N$ input photons on $e^{-iH_b^{\text{1.exc}}t}$ can be used to probe signatures of chaos in $H_s$ within the $m$-magnetization subspace, where $H_b^{\text{1.exc}}$ denotes the projection of $H_b$ onto the one-excitation subspace. The corresponding operator in the one-excitation subspace has a matrix representation in terms of a $K\times K$ matrix. Note that the condition $K \gg N^2$ is not especially restrictive, since it is exactly the condition that is required for boson sampling to be classically intractable~\cite{aaronson2011computational}.

\section{Experimental Details}
  \subsection{Experimental Procedure}
  \label{app: experimental setup}

  \begin{figure*}[ht]
    \centering
    \includegraphics[width=0.9\textwidth]{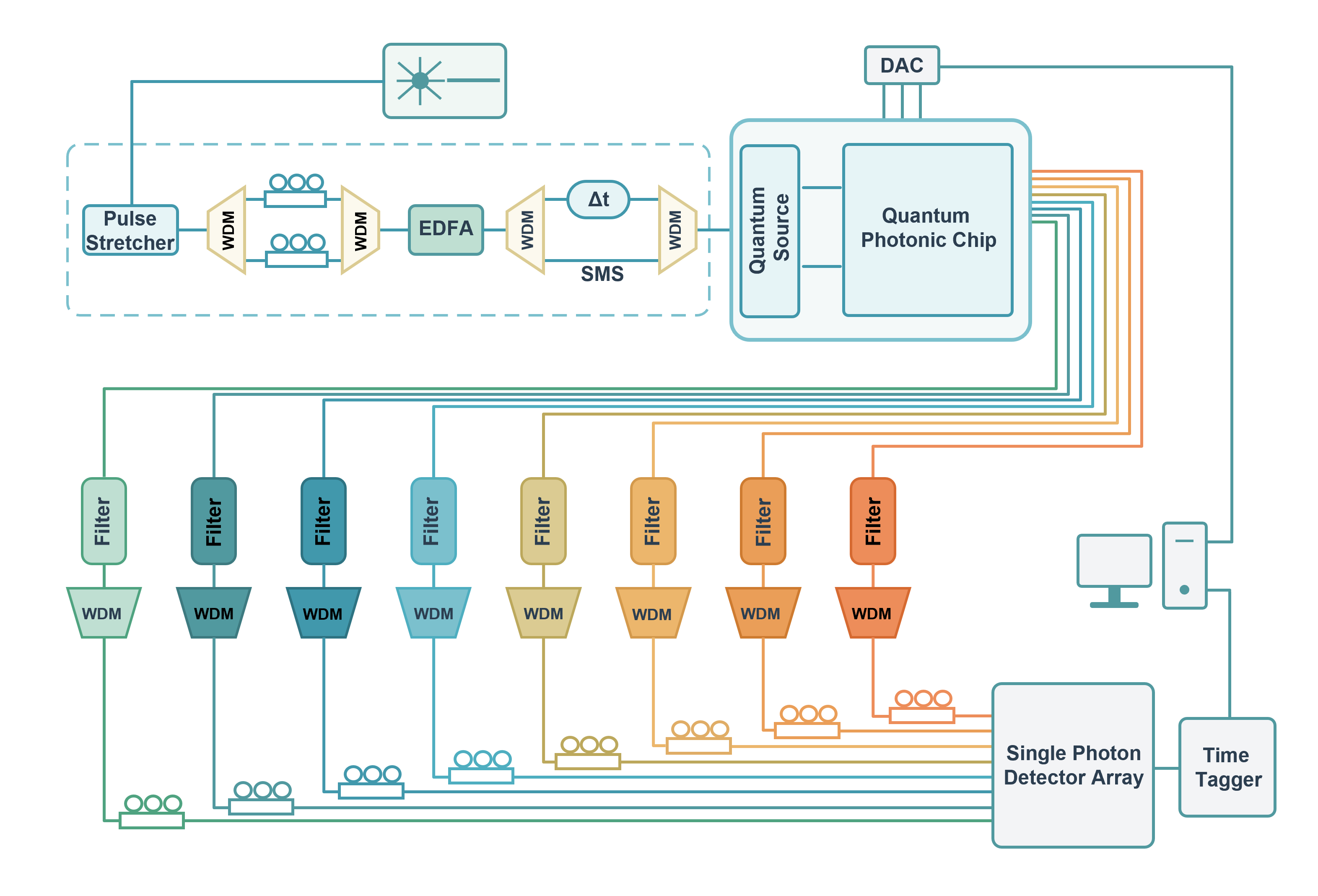}
    \caption{\textbf{The experimental setup for the integrated photonic circuit experiment.} Schematic of the optical source, programmable photonic chip, and photon detection system used for the quantum chaos experiments. }
    \label{fs exp}
  \end{figure*}

  The experimental setup for the quantum chaos study is illustrated in Fig.~\ref{fs exp}. The process involves both an input and output light path, with precise control to ensure high-quality photon generation and detection.

  The pump light is generated using an ultrafast optical clock device and is directed through a pulse stretcher to expand the light's bandwidth to approximately 10 nm. This broad bandwidth is then narrowed by passing the light through a 100-GHz Wavelength Division Multiplexing (WDM) device, which isolates the 1550 nm wavelength with a resolution of 0.5 nm. Following this, a wavelength filter is employed to further refine the linewidth and suppress background noise. The filtered light is then amplified using an Erbium-Doped Fiber Amplifier (EDFA), boosting its power to approximately 100 mW. A second set of WDM devices and wavelength filters is applied to minimize noise in the amplification process while maintaining the desired wavelength's power. To ensure that the optical paths are equalized, tunable fiber delay lines are introduced in one of the channel arms. This adjustment compensates for any discrepancies in the optical path lengths, ensuring synchronized photon arrival. Additionally, polarization controllers are placed in the input path to align the input light mode with the coupling structure. Finally, the prepared pump light is coupled into the photonic chip through fiber grating arrays, which serve as the interface between the external optical system and the on-chip waveguides. The pump light interacts with the integrated photonic structures to generate and manipulate quantum states according to the experimental algorithm.

  After completing the computation within the photonic chip, the output photons are guided through a second set of WDM devices and filters to isolate the desired wavelength and further suppress noise. A polarization controller is placed in the output path to adjust the polarization mode of the outgoing light, ensuring compatibility with the detection system. The filtered photons are then detected using eight-port SNSPDs, which convert the photon signals into electrical signals with high sensitivity and low noise. These electrical signals are processed by a time-tagger, which records the precise timing of each detected photon. The timing data is subsequently transmitted to a computer, which controls the photonic chip’s reconfigurable components. The computer interfaces with the photonic chip via a DAC. The DAC modulates the heating elements on the chip, altering the optical path lengths and reconfiguring the circuit for the next measurement round.

  \subsection{Chip Calibration}
  \label{app: chip calibration}

  The calibration procedure determines the relationship between heater power and adjustable phases for each MZI. We first consider the calibration of the phase shifter placed between two beam splitters. When light enters the top arm of the MZI, the output state $P$ at the two ports can be expressed as:
  \begin{equation}
    P= i e^{i \frac{\theta}{2}}\left[\begin{array}{cc}e^{i \phi} \sin \frac{\theta}{2} & e^{i \phi} \cos \frac{\theta}{2} \\ \cos \frac{\theta}{2} & -\sin \frac{\theta}{2}\end{array}\right] \cdot\left[\begin{array}{l}
        1 \\
        0
      \end{array}\right]=i e^{i \frac{\theta}{2}}\left[\begin{array}{c}
        \sin \frac{\theta}{2} \\
        \cos \frac{\theta}{2}
      \end{array}\right],
  \end{equation}
  where the output power distribution depends on the induced phase change $\theta$. This relationship allows interference fringes to be observed as the phase is varied.

  The phase shift $\theta$ originates from the optical path difference caused by a change in the effective refractive index. Thermal phase shifting is achieved by heating a titanium nitride resistor integrated with the waveguide, resulting in a temperature-induced refractive index change proportional to the heater power:
  \begin{equation}
    \theta=\frac{L}{\lambda} \cdot k \cdot P_\text{heater}+\theta_0,
  \end{equation}
  where $L$ denotes the waveguide length influenced by the heater, $\lambda$ is the wavelength, $\theta_0$ is an intrinsic phase offset due to fabrication imperfections, $P_\text{heater}$ represents the heater power, and $k$ is a proportionality coefficient relating heater power to refractive index modulation.

  Experimentally, we systematically vary the heater power applied to the MZI and measure the resulting output power variation. A fitting procedure is then employed to determine the coefficients $k$ and $\theta_0$. This calibration enables precise control of the phase shifts using heater power settings.

  \subsection{Photon source characterization}
  \label{app: source character}

  The photon source is first characterized under the degenerate spontaneous four-wave mixing process. The spectral properties of the pump laser are measured using an optical spectrum analyzer (OSA). The initial broadband pump exhibits a spectral bandwidth of approximately 1.9 nm. To improve the spectral purity and suppress background noise, the WDMs described in Appendix \ref{app: experimental setup} are used, which reshapes the pump spectrum to a narrower bandwidth of approximately 0.7 nm.

  Two specific wavelength channels, centered at 1553.33 nm and 1546.92 nm, are selected from the broadband pump using the WDM. Through the degenerate SFWM process in the spiral waveguide, correlated photon pairs are generated with identical wavelengths centered at 1550.12 nm.

  \subsection{Sources of Error}
  \label{app: error}

  In this section, we analyze the primary sources of experimental imperfections that affect the performance of the on-chip boson sampling system and the accuracy of quantum chaos diagnostics. These errors mainly arise from photon source non-idealities, imperfect optical transformations, measurement noise, and multiphoton contributions.

  \textbf{Photon source imperfections from SFWM.}
  Photon pairs are generated via SFWM in silicon waveguides. Ideally, the generated two-photon state should be spectrally factorable and indistinguishable across different sources. However, in practice, residual spectral correlations in the joint spectral amplitude lead to partial distinguishability between photons.
  This distinguishability reduces the degree of multiphoton interference, which is essential for boson sampling. As a result, the output probability distribution deviates from the ideal bosonic distribution and tends toward a classical distribution. In the context of quantum chaos characterization, such imperfections can suppress signatures of chaotic behavior, for example by reducing the agreement with the Porter--Thomas distribution and lowering the measured Shannon entropy.

  \textbf{Imperfections in MZI-based unitary transformations.}
  The programmable linear optical network is implemented using cascaded MZIs controlled by thermo-optic phase shifters. Fabrication imperfections and thermal cross-talk introduce deviations from the target unitary transformation.
  Specifically, phase errors arising from heater non-uniformity, limited calibration precision, and thermal drift result in inaccurate setting of the MZI phases. These errors accumulate across the interferometer mesh and lead to a deviation between the implemented unitary and the ideal unitary.
  Such unitary errors distort the output photon distribution and reduce the fidelity of the boson sampling process. In quantum chaos analysis, this may lead to incorrect identification of statistical properties, such as deviations from GOE behavior or artificial broadening of the output distribution.

  \textbf{Noise photons in the measurement system.}
  Noise photons originating from imperfect pump suppression, Raman scattering, and detector dark counts contribute to the measured coincidence events. Although spectral filtering (e.g., WDM and AMZI) is applied to suppress residual pump light, incomplete filtering introduces background counts.
  These noise photons degrade the signal-to-noise ratio and introduce spurious detection events, which bias the reconstructed output distribution. In particular, noise can artificially increase entropy-like measures and obscure the statistical features used to distinguish chaotic and integrable regimes.

  \textbf{Measurement without photon-number-resolving detectors.}
  The lack of photon-number-resolving detection leads to a systematic error associated with unresolved collision events, as analyzed in Appendix~\ref{app: pnr detectors and conditional probabilities}.

    \begin{figure*}[ht]
    \centering
    \includegraphics[width=0.7\textwidth]{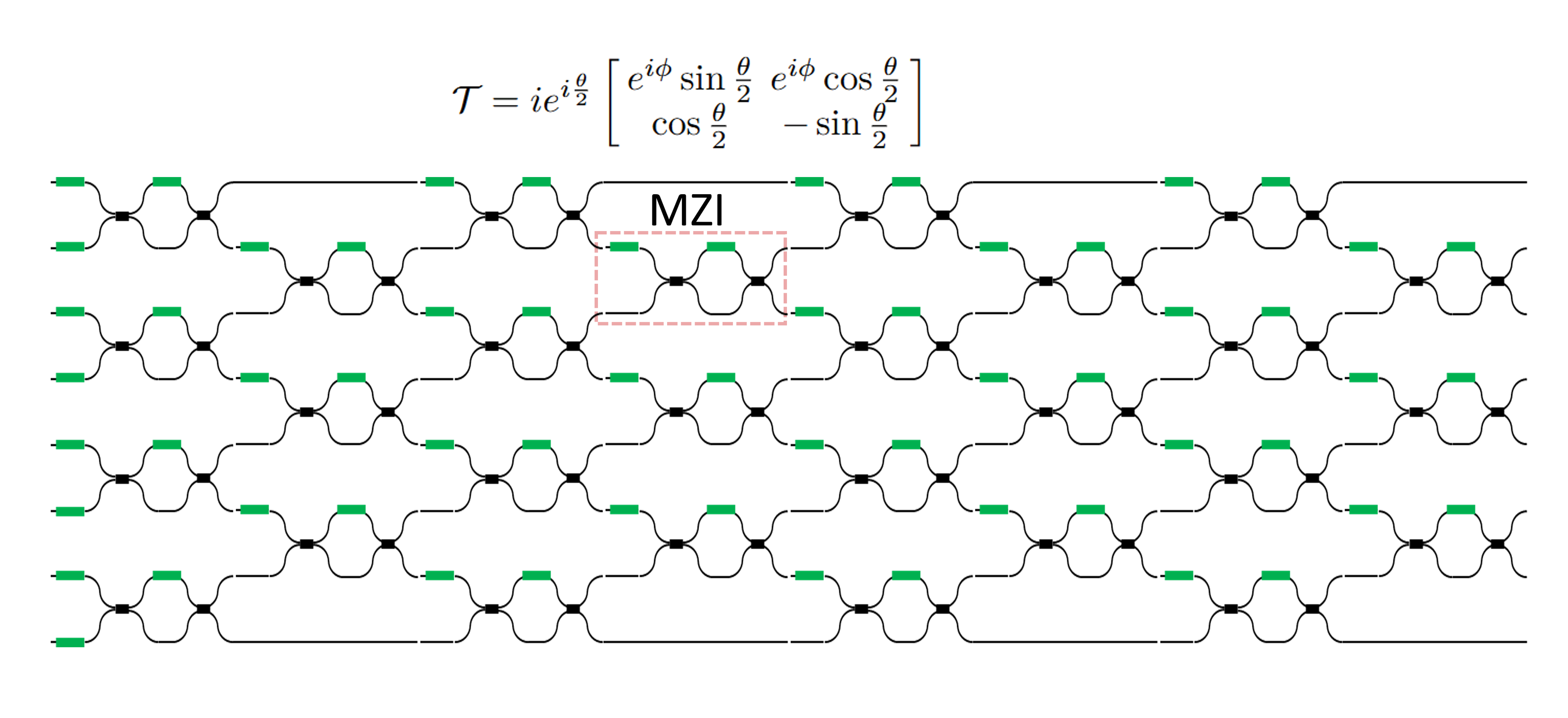}
    \caption{\textbf{Chip decomposition for boson sampling}. Schematic of the programmable MZI mesh architecture used to realize arbitrary unitary transformations for boson sampling experiments.}
    \label{fs1}
  \end{figure*}

  \subsection{Chip decomposition for Boson Sampling}
  \label{app: chip decomposition}

  We use an arrangement of MZIs for constructing universal multiport interferometers, which serve as the
  arbitrary unitaries in the boson sampling algorithm, as shown in Fig.~\ref{fs1}. This design allows each mode to interact with its nearest neighbor, creating a compact and symmetrical structure that optimizes space and efficiency.

  Each fundamental transformation within the MZI unit is governed by two-phase shift parameters, $\theta$ and $\phi$, and is defined as follows,
  \begin{equation}\label{eq: mzi}
    \mathcal{T}=i e^{i \frac{\theta}{2}}\left[\begin{array}{cc}e^{i \phi} \sin \frac{\theta}{2} & e^{i \phi} \cos \frac{\theta}{2} \\ \cos \frac{\theta}{2} & -\sin \frac{\theta}{2}\end{array}\right].
  \end{equation}

  Each MZI operation can be represented by a unitary transformation matrix, denoted by $T_{ij}$, of the form:
  \begin{equation}
    T_{ij}=\left[\begin{array}{cccc}
        1      & \cdots   &          &        \\
        \vdots & a_{i, i} & a_{i, j} &        \\
               & a_{j, i} & a_{j, j} & \vdots \\
               &          & \cdots   & 1
      \end{array}\right]_N.
  \end{equation}
  The parameters of the $T_{i,j}$ matrices determine the values of the beam splitters and phase shifts corresponding to ~\eqref{eq: mzi}. We show that an arbitrary $N \times N$ unitary matrix can be decomposed into $N(N-1)/2$ MZI units arranged in a specific sequence. Repeating this procedure iteratively yields the full target unitary. For additional details on the decomposition method, see our previous work~\cite{zhan2024physics}.

  \section{Theoretical Justification for PT Distribution (Eq.~4) and Shannon Entropy (Eq.~6)}
  \label{app: theoretical justification}

  In this appendix, we provide a theoretical justification for the behavior of the distance $\mathcal{D}$  (Eq.~4) from the PT distribution, and the Shannon entropy (Eq.~6). The analysis establishes the connection between spectral statistics, unitary designs, and boson-sampling underlying the behavior of these two probes.

  \subsection{Spectral Form Factor and Frame Potential}

  For an ensemble of unitary operators \(\{U(t)=e^{-iHt}\}\), the \(k\)-th frame potential
  $\mathcal{F}_k(t)$  provides a measure of how close the ensemble is to forming a unitary \(k\)-design. Smaller values of $\mathcal{F}_k(t)$ indicate closer approximation to Haar-random unitaries, with the Haar ensemble achieving the minimum value \(\mathcal{F}_k(t)= k!\).
  For ensembles of random matrix Hamiltonians (e.g., GOE), it is known that the \(2k\)-point spectral form factor (SFF), denoted  \((\mathcal{R}_{2k})^2\), is related to the frame potential via
  \begin{equation}
    F_k(t) \propto \left[\mathcal{R}_{2k}(t)\right]^2,
  \end{equation}
  up to normalization factors (see Ref.~\cite{cotler2017chaos} ). In chaotic systems, the SFF typically exhibits a dip-ramp-plateau structure: it decreases at early times, reaches a minimum at a characteristic time \(t^*\), and then approaches a plateau at late times.
  This behavior implies that the unitary ensemble \(\{e^{-iHt}\}\) is closest to a unitary design near \(t \approx t^*\), and hence most closely approximates Haar-random behavior at that time.

  \subsection{Connection to Porter--Thomas Distribution}

  Boson sampling with Haar-random unitaries produces output probabilities that follow the Porter--Thomas distribution
  \begin{equation}
    P_{\mathrm{PT}}(p) = D e^{-D p},
  \end{equation}
  where \(D\) is the dimension of the Hilbert space in \eqref{eq:s3}.
  Therefore, if the ensemble \(\{e^{-iHt}\}\) approximates a unitary design at time \(t\), we expect the corresponding boson sampling output probabilities to be close to Porter--Thomas distribution.
  This motivates defining the distance $\mathcal{D}(t)$  (Eq.~4).  Based on the above reasoning, we expect \(\mathcal{D}(t)\) to exhibit a dip near \(t \approx t^*\), corresponding to the minimum of the SFF. At later times, the distance approaches a plateau.
Due to finite system size and limited sampling, this correspondence is expected to hold only approximately rather than exactly.

  \subsection{Shannon Entropy at the Haar Point}
  At times when \(U(t)\) is close to Haar-random, the output probability distribution is approximately uniformly random over the simplex. In this case, one can compute the expected Shannon entropy.
  Let \(p_1, \ldots, p_D\) be the probabilities associated with all output configurations. The Shannon entropy is
  \begin{equation}
    S = -\sum_{i=1}^D p_i \ln p_i \ .
  \end{equation}
  The ensemble average over Haar-random unitaries is given by
  \begin{equation}
    \langle S \rangle_{\mathrm{ens}} = \sum_{i=1}^D \left\langle p_i \ln \left(\frac{1}{p_i}\right) \right\rangle_{\mathrm{ens}} \ .
  \end{equation}
  Using the fact that the distribution of a single probability under the Haar measure is
  \begin{equation}
    P(p) = (D-1)(1-p)^{D-2} \ ,
  \end{equation}
  we obtain
  \begin{equation}
    \langle S \rangle_{\mathrm{ens}} = \sum_{i=1}^D \int_0^1 dp \, p \ln\left(\frac{1}{p}\right) (D-1)(1-p)^{D-2} \ .
  \end{equation}
  Evaluating the integral gives
  \begin{equation}
    \label{appendixEQ: Sens part 1}
    \langle S \rangle_{\mathrm{ens}} = -1 + \sum_{i=1}^D \frac{1}{i} \ .
  \end{equation}
  In the large-\(D\) limit, we can approximate 
  \begin{equation}
    (D-1)(1-p)^{D-2} \approx De^{-Dp} \ ,
  \end{equation}
  corresponding to the Porter--Thomas distribution. Substituting this approximation into the entropy gives
  \begin{align}
    \langle S \rangle_{\mathrm{ens}} & \approx \sum_{i=1}^D \int_0^\infty dp \, p \ln\left(\frac{1}{p}\right) D e^{-Dp} \\
                                     & = -1 + \ln(D) + \gamma,
  \end{align}
  where $\gamma$ is the Euler-Mascheroni constant. Equation~\ref{appendixEQ: Sens part 1} is used in the main text.

  \subsection{Connection to Experimental Observables}
  \label{app: theoretical justification E3}
  In our experiment, the quantity \(S_{\text{avg}}(t)\) defined in Eq.~6 serves as an estimator for \(\langle S \rangle_{\mathrm{ens}}\). By the Central Limit Theorem, we expect that for sufficiently large ensembles,
  \begin{equation}
    S_{\text{avg}}(t) \approx \langle S \rangle_{\mathrm{ens}}.
  \end{equation}

  On the other hand, as the Hilbert-space dimension increases, concentration-of-measure results (e.g., Lévy’s lemma) imply that, when the unitary ensemble is close to Haar random, a typical realization satisfies
  \begin{equation}
    S \approx \langle S \rangle_{\mathrm{ens}},
  \end{equation}
  independently of the particular probabilities \(\{p_i\}_i\) used to calculate \(S\). 
  Hence, for small Hilbert-space dimensions, the estimate of \(\langle S \rangle_{\mathrm{ens}}\) can be improved by increasing the ensemble size. For sufficiently large dimensions, and when the unitary ensemble is close to Haar-random behavior, even a single realization is expected to produce an entropy value close to the ensemble average. In both regimes, the entropy is expected to reach its maximum near the characteristic time \(t^*\), where \(\mathcal{D}(t)\) is minimized and the SFF reaches its dip.

  \subsection{Summary}

  The above analysis provides the following interpretation for chaotic Hamiltonians. The spectral form factor exhibits the characteristic dip--ramp--plateau structure, with a minimum occurring at \(t \approx t^*\). Around this time, the unitary evolution \(e^{-iHt}\) approaches Haar-random behavior.
  As a consequence, the boson-sampling output probabilities become closest to the Porter--Thomas distribution, producing a minimum in the distance \(\mathcal{D}(t)\). At the same time, the Shannon entropy is maximized and approaches its Haar-average value.

 For integrable systems, the spectral form factor does not exhibit the same dip--ramp--plateau structure associated with chaotic dynamics. Correspondingly, the boson-sampling output statistics are not expected to approach Porter--Thomas behavior, and no corresponding minimum in \(\mathcal{D}(t)\) or maximum in \(S_{\text{avg}}(t)\) is expected.

\section{Numerical Scaling of Chaos Probes with System Size}
  \label{app: scaling arguments}

  \begin{figure*}
    \centering
    \includegraphics[width=0.98\linewidth]{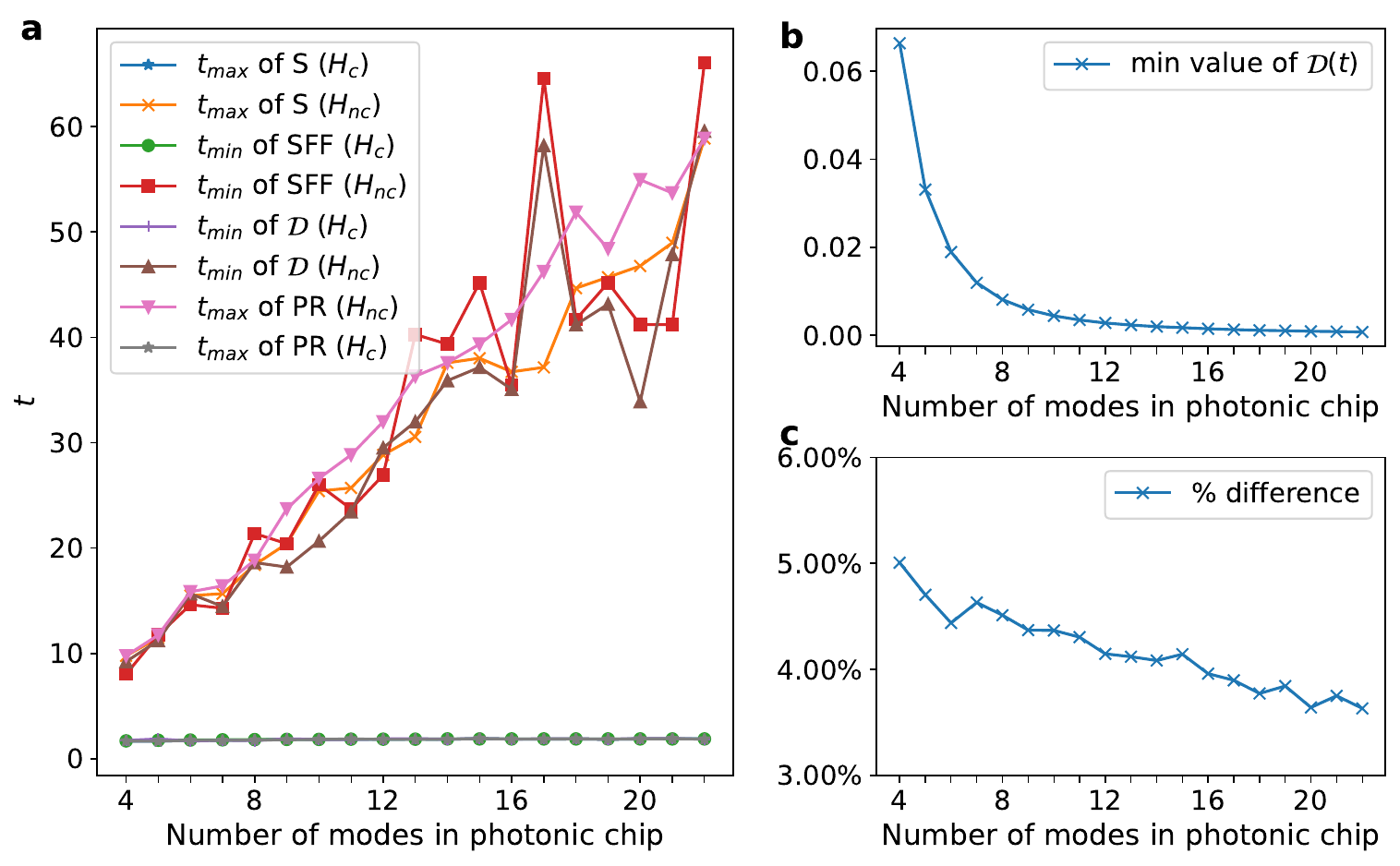}
    \caption{\textbf{Numerical scaling behavior of the chaos probes.} 
      \textbf{a)} For both the chaotic and non-chaotic cases, we plot the time corresponding to the minimum of the SFF, the time where the empirical distribution of probability values is the closest to the PT distribution, the time corresponding to the maximum of the Shannon entropy, and the time where the PR of the OTOCs reaches its maximum value. For the chaotic times, all four times converge and are approximately constant, whereas no such convergence is observed in the non-chaotic case.
      \textbf{b)} For the chaotic case, we show that the empirical distribution of probability values approaches the PT distribution at $t^*$ as the system size increases.
      \textbf{c)} The percentage difference between $\langle S \rangle_{\mathrm{Haar}}$ (Eq.~7) and the $\max_{t} S_{\mathrm{avg}}$ (Eq.~8), obtained from boson-sampling simulations with 2000 realizations. The maximal value of the Shannon Entropy gets closer to $\langle S \rangle_{\mathrm{Haar}}$ as the system size increases.}
    \label{fig: scaling result SFF PT SE}
  \end{figure*}
The proposed probes exhibit increasingly clear signatures as the system size increases, as shown in Fig.~\ref{fig: scaling result SFF PT SE}. Figure~\ref{fig: scaling result SFF PT SE}\textbf{a} compares the times corresponding to the minimum of the SFF, the minimum of $\mathcal{D}(t)$, the maximum of the averaged Shannon entropy, and the maximum of PR for both chaotic and integrable dynamics. In the chaotic case, these four times are approximately the same value, nearly independent of the system size. In contrast, in the integrable case,  these four times are not the same and become more pronounced as the number of modes increases.

Figure~\ref{fig: scaling result SFF PT SE}\textbf{b} shows that the minimum point of $\mathcal{D}$ gets closer to $0$ as the number of modes increases, which is consistent with the reduction in finite size effects.

Figure~\ref{fig: scaling result SFF PT SE}\textbf{c} further shows the percentage difference between $\langle S \rangle_{\mathrm{Haar}}$ (Eq.~7) and the simulated maximal value of the Shannon Entropy, $\max_{t} S_{\mathrm{avg}}(t)$ (Eq.~8), obtained by simulating boson sampling for the chaotic case. The reduction of finite-size effects with increasing system size is consistent with the theoretical analysis presented in Appendix~\ref{app: theoretical justification}.  Taken together, these numerical results indicate that the boson-sampling-based probes become more effective for distinguishing chaotic and integrable dynamics in larger systems.

  \section{OTOC Dynamics in Boson Sampling}
  \label{app: more OTOC behavior}
\begin{figure*}[ht]
    \centering
    \includegraphics[width=0.98\linewidth]{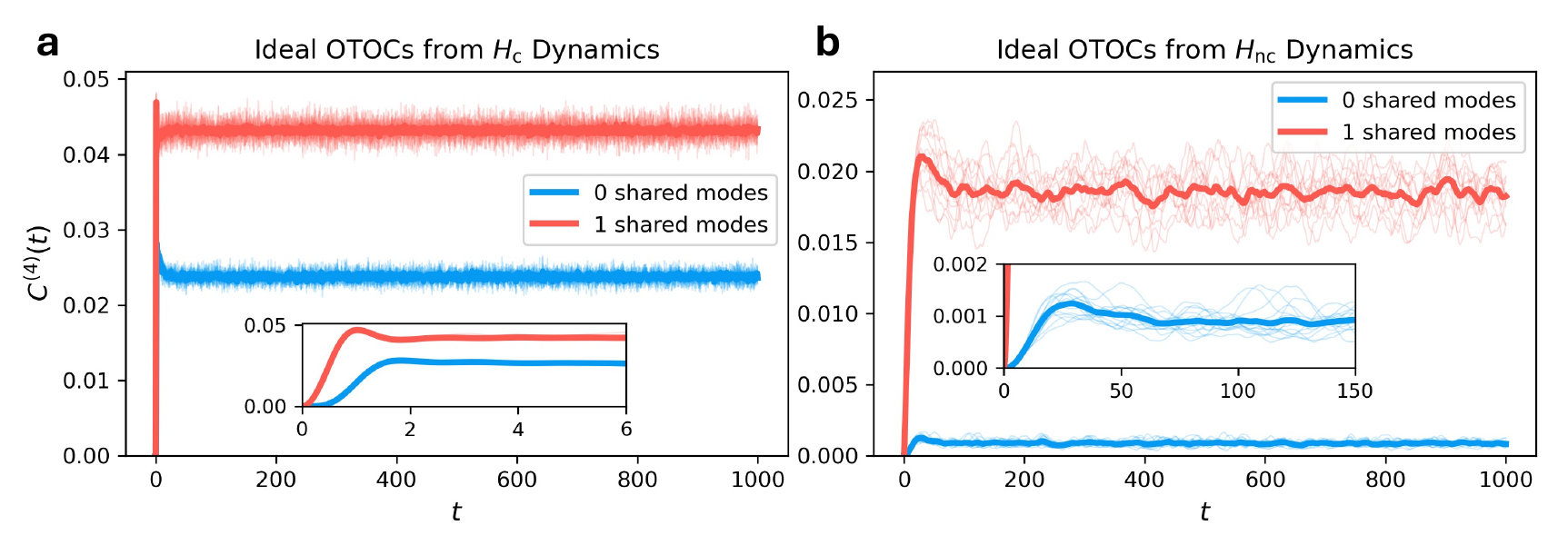}
    \caption{\textbf{Ideal four-point OTOCs for all collision-free output configurations.} \textbf{a)} and \textbf{b)} Ideal four-point OTOCs for all collision-free output configurations under $H_\text{c}$ and $H_\text{nc}$ dynamics, obtained by averaging over 2000 Hamiltonian realizations. Blue (red) curves correspond to configurations with 0 (1) shared occupied modes with the input configuration. Thin curves denote individual configurations, while bold curves represent sector averages. Insets show short-time dynamics.
    }
    \label{fig: ideal_OTOCs}
  \end{figure*}

  As discussed in Section 4.3, the output probabilities in a boson sampling experiment can be mapped to four-point out-of-time-ordered correlators. This correspondence implies that the temporal behavior of the measured output probabilities reflects the scrambling properties of the underlying unitary dynamics. Figure~\ref{fig: ideal_OTOCs} shows the ideal OTOCs for all collision-free output configurations under chaotic and integrable dynamics, grouped by their overlap with the input state. Generally, the OTOC dynamics can be characterized by an initial growth phase up to a scrambling time $t^\star$, after which the correlator saturates and fluctuates around a steady-state value~\cite{wisniacki_PRE2019_OTOC}.
  In this appendix, we analyze both the short-time and long-time regimes of these OTOC-equivalent quantities and show how they can be used as additional diagnostics of chaotic versus integrable dynamics.

  \subsection{Short-Time Behavior}  \label{app:Short-Time Behavior}

  At short times, the four-point OTOC $C^{(4)}_{i,j,r,s}(t)$
  can be expanded perturbatively in time. For a two-photon input state $\hat{a}_i^\dagger \hat{a}_j^\dagger |0\rangle$ and output modes $r$ and $s$, the OTOC-equivalent observable is
  \begin{equation}
    C^{(4)}_{i,j,r,s}(t) = \langle 0 | \hat{C}^\dagger_{i,j,r,s}(t)\, \hat{C}_{i,j,r,s}(t) |0\rangle,
    \label{eqn: 4-point OTOC Appendix}
  \end{equation}
  where $\hat{C}_{i,j,r,s}(t)=[\hat{a}_i^\dagger(t)\hat{a}_j^\dagger(t), \hat{a}_r \hat{a}_s]$ and the Heisenberg-evolved creation operator is given by $\hat{a}_i^\dagger(t) = \sum_k U_{ik}(t)\hat{a}_k^\dagger$. Substituting the Heisenberg-evolved operators into the commutator gives
  \begin{equation}
    \hat{C}_{i,j,r,s}(t) = \sum_{k,l} U_{ik}(t)U_{jl}(t)[\hat{a}_k^\dagger\hat{a}_l^\dagger, \hat{a}_r \hat{a}_s].
  \end{equation}
  Using the bosonic commutation relations gives the usual two-photon bosonic interference term,
  \begin{equation}
    C^{(4)}_{i,j,r,s}(t) = \left|U_{ir}(t)U_{js}(t) + U_{is}(t)U_{jr}(t)\right|^2,
    \label{eqn: 4-point OTOC 2}
  \end{equation}
  up to the normalization convention used for collision-free output probabilities. Thus, the short-time behavior of the OTOC is determined by the leading powers of the single-particle transition amplitudes $U_{ab}(t)$. Expanding $U(t)=e^{-iHt}$, we have $U_{ab}=\delta_{ab}-itH_{ab}+\mathcal{O}(t^2)$. Therefore, diagonal amplitudes are of order unity at short times, $U_{aa}(t) = 1+\mathcal{O}(t)$, while off-diagonal transition amplitudes are of order $t$, $U_{ab}(t)=\mathcal{O}(t)$ with $a\neq b$.

  The power law then follows from the number of off-diagonal transitions needed to connect the input configuration $(i,j)$ to the output configuration $(r,s)$. If the final configuration shares one occupied mode with the input configuration, for example $r=i$ and $s \neq j$, then $U_{ir}(t)U_{js}(t)=U_{ii}(t)U_{js}(t)=\mathcal{O}(1)\mathcal{O}(t)=\mathcal{O}(t)$, while the exchange term $U_{is}(t)U_{jr}(t)$ is at most of order $t^2$. Hence, the leading contribution to the amplitude is of order $t$, and
  \begin{equation}
    C^{(4)}_{i,j,r,s}(t) \propto t^2 \ ,
  \end{equation}
  for output configurations sharing one occupied mode with the input configuration. If there is no overlap between the input and output configurations, then all four single-particle amplitudes appearing in~\eqref{eqn: 4-point OTOC 2} are off-diagonal, $U_{ir}(t),U_{js}(t),U_{is}(t),U_{jr}(t)=\mathcal{O}(t)$. Both bosonic paths in~\eqref{eqn: 4-point OTOC 2} therefore contribute at order $t^2$, so the probability scales as
  \begin{equation}
    C^{(4)}_{i,j,r,s}(t) \propto t^4. \
  \end{equation}
  More generally, the short-time scaling exponent is determined by the minimum number of off-diagonal transitions required to connect the input and output configurations. Configurations sharing occupied modes with the input state therefore dominate at early times, reflecting perturbative memory retention of the initial configuration.

\begin{figure*}
    \centering
    \includegraphics[width=0.99\linewidth]{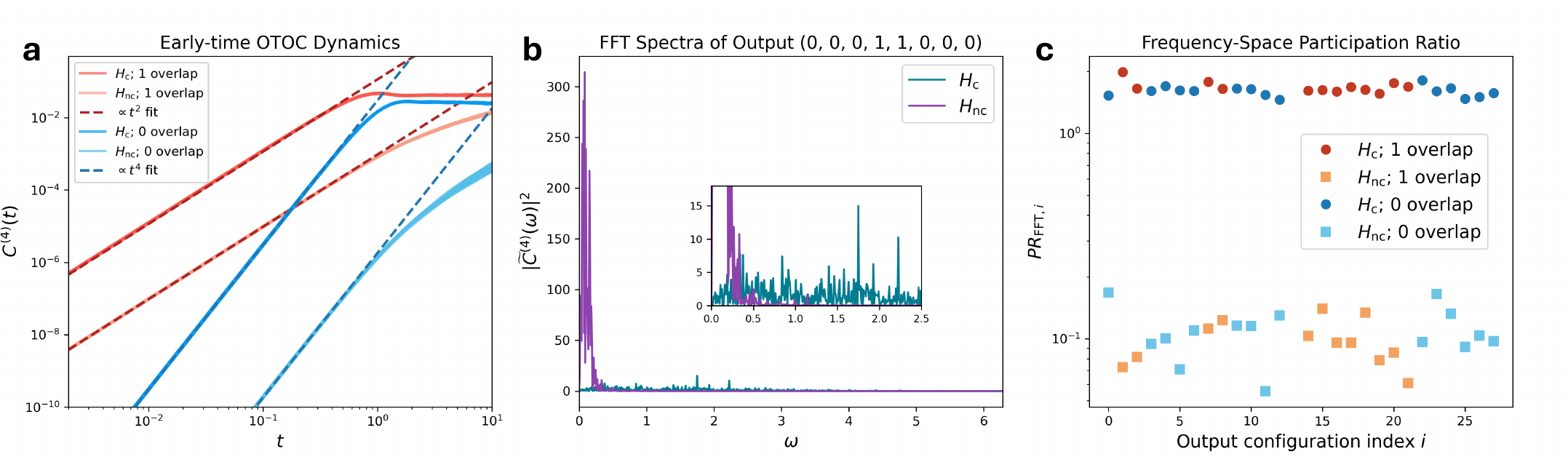}
    \caption{\textbf{Additional signatures of quantum chaos from OTOC-equivalent observables.} \textbf{a)} Early-time OTOCs exhibit $t^2$ scaling for output configurations sharing one occupied mode with the input configuration and $t^4$ scaling for configurations with no shared occupied modes; dashed lines denote the fitted power-law scalings. All output configurations are plotted and grouped according to the number of shared occupied modes with the initial configuration. \textbf{b)} FFT power spectra $\left|\widetilde{C}^{(4)}(\omega)\right|^2$ of the late-time OTOC fluctuations for the output configuration $\ket{0_1,0_2,0_3,1_4,1_5,0_6,0_7,0_8}$ are shown for $H_{\rm{c}}$ (teal) and $H_{\rm{nc}}$ (purple) dynamics. Before computing the FFT, the OTOCs from $t=300$ to $t=1000$ are normalized by their temporal mean values and shifted by subtracting unity.  \textbf{c)} Frequency-space participation ratios ${\rm PR_{\rm FFT}}$ of the Fourier-transformed OTOCs are shown for all output configurations except the initial configuration. Before computing the ${\rm PR_{\rm FFT}}$, the FFT power spectra in \textbf{b)} are normalized such that the integrals are unity. Circular markers correspond to GOE dynamics, while square markers correspond to Poisson dynamics.}
    \label{fig: extra_OTOC_result}
  \end{figure*}

  This explains the two short-time power laws shown in  Fig.~\ref{fig: extra_OTOC_result}\textbf{a}: configurations with one shared occupied mode scale as $t^2$, while configurations with no shared occupied modes scale as $t^4$. The scaling is determined by the perturbative structure of the single-particle transition amplitudes and therefore does not by itself distinguish chaotic from integrable dynamics. The difference between the two regimes appears only in the prefactors, with integrable (Poissonian) dynamics typically yielding smaller amplitudes, and in the long-time behavior shown below. Similar short-time power-law behavior has been observed in other systems, such as spin chains with random fields~\cite{wisniacki_PRE2019_OTOC}, confirming that this regime reflects general operator growth rather than chaos-specific features.

  \subsection{Long-Time Behavior and Fluctuations} \label{app:Long-Time Behavior}

  At longer times, the OTOC approaches a steady-state value and exhibits fluctuations around it. The nature of these fluctuations provides additional information about the underlying dynamics.
  Fig.~\ref{fig: extra_OTOC_result}\textbf{b} shows the Fourier power spectrum of the time-dependent OTOC for a representative output configuration. In the chaotic case, the spectrum contains contributions from a broader range of frequencies, reflecting more complex temporal fluctuations and stronger mixing in Hilbert space. By contrast, the integrable spectrum is more localized, consistent with more regular and constrained dynamics.

  \subsection{Frequency-Space Delocalization}

  To quantify the spread of the OTOC in frequency space, we calculate the participation ratio of the Fourier-transformed correlator:
  \begin{equation}
    \mathrm{PR}_{\mathrm{FFT}} = \left[\int_0^\infty d\omega\, |\tilde{C}(\omega)|^4 \right]^{-1}, \quad \int_0^\infty d\omega\, |\tilde{C}(\omega)|^2=1,
    \label{eqn: FFT IPR}
  \end{equation}
  where \(\tilde{C}(\omega)\) is the normalized Fourier-transformed OTOC of a specific output configuration at late times.
  This quantity measures the effective number of frequency components contributing to the OTOC fluctuations: small values correspond to spectra concentrated around a few dominant frequencies, while large values indicate more delocalized spectra.

  Fig.~\ref{fig: extra_OTOC_result}\textbf{c} shows that \(\mathrm{PR}_{\mathrm{FFT}}\) is systematically larger in the chaotic case than in the integrable case across all collision-free output configurations other than the initial state. This provides a quantitative measure of the enhanced complexity of temporal fluctuations in chaotic dynamics. Unlike the short-time scaling behavior, which is strongly organized by overlap sectors, the separation in \(\mathrm{PR}_{\mathrm{FFT}}\) persists broadly across output configurations, indicating that the long-time frequency-space complexity is governed more directly by the underlying dynamics.
  This long-time fluctuation behavior of the OTOC has also been observed in spin chains~\cite{wisniacki_PRE2019_OTOC, Omanakuttan_Chinni_Poggi_PRA_2023}.

 \subsection{Interpretation and Additional Remarks}

  Taken together, the results presented in this appendix provide a consistent picture of the OTOC dynamics in boson sampling. The short-time behavior is governed by the perturbative structure of the single-particle transition amplitudes and reflects general operator growth rather than chaos-specific behavior. In particular, the growth follows overlap-dependent power-law scaling, and configurations sharing more occupied modes with the input state dominate at early times.

  At longer times, the OTOC temporal fluctuations and their frequency content can be used to provide qualitative and quantitative diagnostics of chaos. Chaotic dynamics exhibit broader frequency spectra and larger \(\mathrm{PR}_{\mathrm{FFT}}\), consistent with enhanced scrambling and delocalization driven by chaotic evolution. These long-time fluctuations, therefore, provide a complementary probe of scrambling beyond the early-time growth regime.

  These observations complement the diagnostics presented in Section~4 and provide additional signatures distinguishing chaotic and integrable dynamics in the present system.  Extending the analysis to larger system sizes would be an interesting direction for future work.

\end{appendices}

\newpage

\bibliography{sn-bibliography.bib}

@article{Victor_quantum_metamorphism,
  title     = {Quantum metamorphism},
  author    = {Bastidas, V. M. and Estarellas, M. P. and Osada, T. and Nemoto, Kae and Munro, W. J.},
  journal   = {Phys. Rev. B},
  volume    = {102},
  issue     = {22},
  pages     = {224307},
  numpages  = {10},
  year      = {2020},
  month     = {Dec},
  publisher = {American Physical Society},
  doi       = {10.1103/PhysRevB.102.224307},
}

@article{participation_ratio_OTOC_lea,
  title     = {Timescales in the quench dynamics of many-body quantum systems: Participation ratio versus out-of-time ordered correlator},
  author    = {Borgonovi, Fausto and Izrailev, Felix M. and Santos, Lea F.},
  journal   = {Phys. Rev. E},
  volume    = {99},
  issue     = {5},
  pages     = {052143},
  numpages  = {11},
  year      = {2019},
  month     = {May},
  publisher = {American Physical Society},
  doi       = {10.1103/PhysRevE.99.052143}
}

@article{zhu2025gbs,
  title     = {{GBS}-Assisted Quantum Unsupervised Machine Learning on a Universal Programmable Integrated Quantum Chip},
  author    = {Zhu, Huihui and Luo, Wei and Yan, Rudai and Ren, Chao and Guo, Jia and Zhao, Zichao and Ma, Haoran and Chen, Tian and Gao, Feng and Kwek, Leong Chuan and others},
  journal   = {Research},
  volume    = {8},
  pages     = {1006},
  year      = {2025},
  publisher = {AAAS}
}

@article{zhan2024physics,
  title     = {Physics-Aware Analytic-Gradient Training of Photonic Neural Networks},
  author    = {Zhan, Yuancheng and Zhang, Hui and Lin, Hexiang and Chin, Lip Ket and Cai, Hong and Karim, Muhammad Faeyz and Poenar, Daniel Puiu and Jiang, Xudong and Mak, Man-Wai and Kwek, Leong Chuan and others},
  journal   = {Laser \& Photonics Reviews},
  volume    = {18},
  number    = {4},
  pages     = {2300445},
  year      = {2024},
  publisher = {Wiley Online Library}
}

@article{zhang2022resource,
  title     = {Resource-efficient high-dimensional subspace teleportation with a quantum autoencoder},
  author    = {Zhang, Hui and Wan, Lingxiao and Haug, Tobias and Mok, Wai-Keong and Paesani, Stefano and Shi, Yuzhi and Cai, Hong and Chin, Lip Ket and Karim, Muhammad Faeyz and Xiao, Limin and others},
  journal   = {Science advances},
  volume    = {8},
  number    = {40},
  pages     = {eabn9783},
  year      = {2022},
  publisher = {American Association for the Advancement of Science}
}

@article{zhang2021optical,
  title     = {An optical neural chip for implementing complex-valued neural network},
  author    = {Zhang, Hui and Gu, Mile and Jiang, XD and Thompson, Jayne and Cai, Hong and Paesani, Stefano and Santagati, Raffaele and Laing, Anthony and Zhang, Y and Yung, Man-Hong and others},
  journal   = {Nature communications},
  volume    = {12},
  number    = {1},
  pages     = {457},
  year      = {2021},
  publisher = {Nature Publishing Group UK London}
}

@article{scaling_rm_PRA_1991,
  title     = {Scaling laws of the additive random-matrix model},
  author    = {Lenz, Georg and \ifmmode \dot{Z}\else \.{Z}yczkowski, Karol and Saher, Dirk},
  journal   = {Phys. Rev. A},
  volume    = {44},
  issue     = {12},
  pages     = {8043--8050},
  year      = {1991},
  month     = {Dec},
  publisher = {American Physical Society},
  doi       = {10.1103/PhysRevA.44.8043}
}

@article{kulkarni_photonic_PRA_2025,
  title     = {Photonic simulation of localization phenomena using boson sampling},
  author    = {Kulkarni, Anuprita V. and Tiwari, Vatsana and Sharma, Auditya and Raina, Ankur},
  journal   = {Phys. Rev. A},
  volume    = {111},
  issue     = {3},
  pages     = {032622},
  numpages  = {13},
  year      = {2025},
  month     = {Mar},
  publisher = {American Physical Society},
  doi       = {10.1103/PhysRevA.111.032622}
}

@article{hoch2025quantum,
  title     = {Quantum machine learning with Adaptive Boson Sampling via post-selection},
  author    = {Hoch, Francesco and Caruccio, Eugenio and Rodari, Giovanni and Francalanci, Tommaso and Suprano, Alessia and Giordani, Taira and Carvacho, Gonzalo and Spagnolo, Nicol{\`o} and Koudia, Seid and Proietti, Massimiliano and others},
  journal   = {Nature Communications},
  volume    = {16},
  number    = {1},
  pages     = {902},
  year      = {2025},
  publisher = {Nature Publishing Group UK London}
}

@article{deng_solving_graph_PRL_2023,
  title     = {Solving Graph Problems Using Gaussian Boson Sampling},
  author    = {Deng, Yu-Hao and Gong, Si-Qiu and Gu, Yi-Chao and Zhang, Zhi-Jiong and Liu, Hua-Liang and Su, Hao and Tang, Hao-Yang and Xu, Jia-Min and Jia, Meng-Hao and Chen, Ming-Cheng and Zhong, Han-Sen and Wang, Hui and Yan, Jiarong and Hu, Yi and Huang, Jia and Zhang, Wei-Jun and Li, Hao and Jiang, Xiao and You, Lixing and Wang, Zhen and Li, Li and Liu, Nai-Le and Lu, Chao-Yang and Pan, Jian-Wei},
  journal   = {Phys. Rev. Lett.},
  volume    = {130},
  issue     = {19},
  pages     = {190601},
  numpages  = {7},
  year      = {2023},
  month     = {May},
  publisher = {American Physical Society},
  doi       = {10.1103/PhysRevLett.130.190601}
}

@article{madsen2022quantum,
  title     = {Quantum computational advantage with a programmable photonic processor},
  author    = {Madsen, Lars S and Laudenbach, Fabian and Askarani, Mohsen Falamarzi and Rortais, Fabien and Vincent, Trevor and Bulmer, Jacob FF and Miatto, Filippo M and Neuhaus, Leonhard and Helt, Lukas G and Collins, Matthew J and others},
  journal   = {Nature},
  volume    = {606},
  number    = {7912},
  pages     = {75--81},
  year      = {2022},
  publisher = {Nature Publishing Group UK London}
}

@article{spring2013boson,
  title     = {Boson sampling on a photonic chip},
  author    = {Spring, Justin B and Metcalf, Benjamin J and Humphreys, Peter C and Kolthammer, W Steven and Jin, Xian-Min and Barbieri, Marco and Datta, Animesh and Thomas-Peter, Nicholas and Langford, Nathan K and Kundys, Dmytro and others},
  journal   = {Science},
  volume    = {339},
  number    = {6121},
  pages     = {798--801},
  year      = {2013},
  publisher = {American Association for the Advancement of Science}
}

@article{flamini2018photonic,
  title     = {Photonic quantum information processing: a review},
  author    = {Flamini, Fulvio and Spagnolo, Nicolo and Sciarrino, Fabio},
  journal   = {Reports on Progress in Physics},
  volume    = {82},
  number    = {1},
  pages     = {016001},
  year      = {2018},
  publisher = {IOP Publishing}
}

@article{wang2020integrated,
  title     = {Integrated photonic quantum technologies},
  author    = {Wang, Jianwei and Sciarrino, Fabio and Laing, Anthony and Thompson, Mark G},
  journal   = {Nature photonics},
  volume    = {14},
  number    = {5},
  pages     = {273--284},
  year      = {2020},
  publisher = {Nature Publishing Group UK London}
}

@article{lenzini2018integrated,
  title     = {Integrated photonic platform for quantum information with continuous variables},
  author    = {Lenzini, Francesco and Janousek, Jiri and Thearle, Oliver and Villa, Matteo and Haylock, Ben and Kasture, Sachin and Cui, Liang and Phan, Hoang-Phuong and Dao, Dzung Viet and Yonezawa, Hidehiro and others},
  journal   = {Science advances},
  volume    = {4},
  number    = {12},
  pages     = {eaat9331},
  year      = {2018},
  publisher = {American Association for the Advancement of Science}
}

@book{nielsen2010quantum,
  title     = {Quantum computation and quantum information},
  author    = {Nielsen, Michael A and Chuang, Isaac L},
  year      = {2010},
  publisher = {Cambridge university press}
}

@article{brod2019photonic,
  title     = {Photonic implementation of boson sampling: a review},
  author    = {Brod, Daniel J and Galv{\~a}o, Ernesto F and Crespi, Andrea and Osellame, Roberto and Spagnolo, Nicol{\`o} and Sciarrino, Fabio},
  journal   = {Advanced Photonics},
  volume    = {1},
  number    = {3},
  pages     = {034001--034001},
  year      = {2019},
  publisher = {Society of Photo-Optical Instrumentation Engineers}
}

@article{lemos2012experimental,
  title     = {Experimental observation of quantum chaos in a beam of light},
  author    = {Lemos, Gabriela B and Gomes, Rafael M and Walborn, Stephen P and Souto Ribeiro, Paulo H and Toscano, Fabricio},
  journal   = {Nature communications},
  volume    = {3},
  number    = {1},
  pages     = {1211},
  year      = {2012},
  publisher = {Nature Publishing Group UK London}
}

@article{Ying_PRA_2025,
  title     = {Many-body quantum chaos, localization, and multiphoton entanglement in optical synthetic frequency dimension},
  author    = {Wang, Junlin and Wang, Luojia and Ma, Jinlou and Yang, Ang and Yuan, Luqi and Ying, Lei},
  journal   = {Phys. Rev. Appl.},
  volume    = {23},
  issue     = {3},
  pages     = {034076},
  numpages  = {17},
  year      = {2025},
  month     = {Mar},
  publisher = {American Physical Society},
  doi       = {10.1103/PhysRevApplied.23.034076}
}

@article{yu2024neumann,
  title     = {A von-Neumann-like photonic processor and its application in studying quantum signature of chaos},
  author    = {Yu, Shang and Liu, Wei and Tao, Si-Jing and Li, Zhi-Peng and Wang, Yi-Tao and Zhong, Zhi-Peng and Patel, Raj B and Meng, Yu and Yang, Yuan-Ze and Wang, Zhao-An and others},
  journal   = {Light: Science \& Applications},
  volume    = {13},
  number    = {1},
  pages     = {74},
  year      = {2024},
  publisher = {Nature Publishing Group UK London}
}

@article{Omanakuttan_Chinni_Poggi_PRA_2023,
  title     = {Scrambling and quantum chaos indicators from long-time properties of operator distributions},
  author    = {Omanakuttan, Sivaprasad and Chinni, Karthik and Blocher, Philip Daniel and Poggi, Pablo M.},
  journal   = {Phys. Rev. A},
  volume    = {107},
  issue     = {3},
  pages     = {032418},
  numpages  = {15},
  year      = {2023},
  month     = {Mar},
  publisher = {American Physical Society},
  doi       = {10.1103/PhysRevA.107.032418}
}

@article{Carlos_Santos_Hirsch_PRL_2019,
  title     = {Quantum and Classical Lyapunov Exponents in Atom-Field Interaction Systems},
  author    = {Ch\'avez-Carlos, Jorge and L\'opez-del-Carpio, B. and Bastarrachea-Magnani, Miguel A. and Str\'ansk\'y, Pavel and Lerma-Hern\'andez, Sergio and Santos, Lea F. and Hirsch, Jorge G.},
  journal   = {Phys. Rev. Lett.},
  volume    = {122},
  issue     = {2},
  pages     = {024101},
  numpages  = {7},
  year      = {2019},
  month     = {Jan},
  publisher = {American Physical Society},
  doi       = {10.1103/PhysRevLett.122.024101}
}

@article{garttner2017measuring,
  author   = {G{\"a}rttner, Martin
              and Bohnet, Justin G.
              and Safavi-Naini, Arghavan
              and Wall, Michael L.
              and Bollinger, John J.
              and Rey, Ana Maria},
  title    = {Measuring out-of-time-order correlations and multiple quantum spectra in a trapped-ion quantum magnet},
  journal  = {Nature Physics},
  year     = {2017},
  month    = {Aug},
  day      = {01},
  volume   = {13},
  number   = {8},
  pages    = {781-786},
  issn     = {1745-2481},
  doi      = {10.1038/nphys4119}
}

@article{li2017measuring,
  title     = {Measuring Out-of-Time-Order Correlators on a Nuclear Magnetic Resonance Quantum Simulator},
  author    = {Li, Jun and Fan, Ruihua and Wang, Hengyan and Ye, Bingtian and Zeng, Bei and Zhai, Hui and Peng, Xinhua and Du, Jiangfeng},
  journal   = {Phys. Rev. X},
  volume    = {7},
  issue     = {3},
  pages     = {031011},
  numpages  = {12},
  year      = {2017},
  month     = {Jul},
  publisher = {American Physical Society},
  doi       = {10.1103/PhysRevX.7.031011}
}

@article{xiao2021information,
  author   = {Xiao Mi  and Pedram Roushan  and Chris Quintana  and Salvatore Mandrà  and Jeffrey Marshall  and Charles Neill  and Frank Arute  and Kunal Arya  and Juan Atalaya  and Ryan Babbush  and Joseph C. Bardin  and Rami Barends  and Joao Basso  and Andreas Bengtsson  and Sergio Boixo  and Alexandre Bourassa  and Michael Broughton  and Bob B. Buckley  and David A. Buell  and Brian Burkett  and Nicholas Bushnell  and Zijun Chen  and Benjamin Chiaro  and Roberto Collins  and William Courtney  and Sean Demura  and Alan R. Derk  and Andrew Dunsworth  and Daniel Eppens  and Catherine Erickson  and Edward Farhi  and Austin G. Fowler  and Brooks Foxen  and Craig Gidney  and Marissa Giustina  and Jonathan A. Gross  and Matthew P. Harrigan  and Sean D. Harrington  and Jeremy Hilton  and Alan Ho  and Sabrina Hong  and Trent Huang  and William J. Huggins  and L. B. Ioffe  and Sergei V. Isakov  and Evan Jeffrey  and Zhang Jiang  and Cody Jones  and Dvir Kafri  and Julian Kelly  and Seon Kim  and Alexei Kitaev  and Paul V. Klimov  and Alexander N. Korotkov  and Fedor Kostritsa  and David Landhuis  and Pavel Laptev  and Erik Lucero  and Orion Martin  and Jarrod R. McClean  and Trevor McCourt  and Matt McEwen  and Anthony Megrant  and Kevin C. Miao  and Masoud Mohseni  and Shirin Montazeri  and Wojciech Mruczkiewicz  and Josh Mutus  and Ofer Naaman  and Matthew Neeley  and Michael Newman  and Murphy Yuezhen Niu  and Thomas E. O’Brien  and Alex Opremcak  and Eric Ostby  and Balint Pato  and Andre Petukhov  and Nicholas Redd  and Nicholas C. Rubin  and Daniel Sank  and Kevin J. Satzinger  and Vladimir Shvarts  and Doug Strain  and Marco Szalay  and Matthew D. Trevithick  and Benjamin Villalonga  and Theodore White  and Z. Jamie Yao  and Ping Yeh  and Adam Zalcman  and Hartmut Neven  and Igor Aleiner  and Kostyantyn Kechedzhi  and Vadim Smelyanskiy  and Yu Chen },
  title    = {Information scrambling in quantum circuits},
  journal  = {Science},
  volume   = {374},
  number   = {6574},
  pages    = {1479-1483},
  year     = {2021},
  doi      = {10.1126/science.abg5029},
  eprint   = {https://www.science.org/doi/pdf/10.1126/science.abg5029}
}

@article{clements2016optimal,
  author    = {William R. Clements and Peter C. Humphreys and Benjamin J. Metcalf and W. Steven Kolthammer and Ian A. Walmsley},
  journal   = {Optica},
  number    = {12},
  pages     = {1460--1465},
  publisher = {Optica Publishing Group},
  title     = {Optimal design for universal multiport interferometers},
  volume    = {3},
  month     = {Dec},
  year      = {2016},
  doi       = {10.1364/OPTICA.3.001460}
}

@article{boxio2018characterizing,
  author   = {Boixo, Sergio
              and Isakov, Sergei V.
              and Smelyanskiy, Vadim N.
              and Babbush, Ryan
              and Ding, Nan
              and Jiang, Zhang
              and Bremner, Michael J.
              and Martinis, John M.
              and Neven, Hartmut},
  title    = {Characterizing quantum supremacy in near-term devices},
  journal  = {Nature Physics},
  year     = {2018},
  month    = {Jun},
  day      = {01},
  volume   = {14},
  number   = {6},
  pages    = {595-600},
  issn     = {1745-2481},
  doi      = {10.1038/s41567-018-0124-x}
}

@article{maldacena2016a,
  author   = {Maldacena, Juan
              and Shenker, Stephen H.
              and Stanford, Douglas},
  title    = {A bound on chaos},
  journal  = {Journal of High Energy Physics},
  year     = {2016},
  month    = {Aug},
  day      = {17},
  volume   = {2016},
  number   = {8},
  pages    = {106},
  issn     = {1029-8479},
  doi      = {10.1007/JHEP08(2016)106}
}

@misc{aaronson2011computational,
  title     = {The computational complexity of linear optics},
  author    = {Aaronson, Scott and Arkhipov, Alex},
  note      = {Proceedings of the forty-third annual ACM symposium on Theory of computing},
  pages     = {333--342},
  year      = {2011}
}

@article{spagnolo2013general,
  title     = {General Rules for Bosonic Bunching in Multimode Interferometers},
  author    = {Spagnolo, Nicol\`o and Vitelli, Chiara and Sansoni, Linda and Maiorino, Enrico and Mataloni, Paolo and Sciarrino, Fabio and Brod, Daniel J. and Galv\~ao, Ernesto F. and Crespi, Andrea and Ramponi, Roberta and Osellame, Roberto},
  journal   = {Phys. Rev. Lett.},
  volume    = {111},
  issue     = {13},
  pages     = {130503},
  numpages  = {5},
  year      = {2013},
  month     = {Sep},
  publisher = {American Physical Society},
  doi       = {10.1103/PhysRevLett.111.130503}
}

@article{huh2015boson,
  author   = {Huh, Joonsuk
              and Guerreschi, Gian Giacomo
              and Peropadre, Borja
              and McClean, Jarrod R.
              and Aspuru-Guzik, Al{\'a}n},
  title    = {Boson sampling for molecular vibronic spectra},
  journal  = {Nature Photonics},
  year     = {2015},
  month    = {Sep},
  day      = {01},
  volume   = {9},
  number   = {9},
  pages    = {615-620},
  issn     = {1749-4893},
  doi      = {10.1038/nphoton.2015.153}
}

@article{bastidas2025equilibration,
  title     = {Equilibration of noninteracting photons and quantum signatures of chaos},
  author    = {Bastidas, V. M. and Nourse, H. L. and Sakurai, A. and Hayashi, A. and Nishio, S. and Nemoto, Kae and Munro, W. J.},
  journal   = {Phys. Rev. B},
  volume    = {112},
  issue     = {13},
  pages     = {134304},
  numpages  = {17},
  year      = {2025},
  month     = {Oct},
  publisher = {American Physical Society},
  doi       = {10.1103/tmw1-vry7},
}

@article{garcia2022out,
  title   = {Out-of-time-order correlators and quantum chaos},
  author  = {Garc{\'\i}a-Mata, Ignacio and Jalabert, Rodolfo A and Wisniacki, Diego A},
  journal = {arXiv preprint arXiv:2209.07965},
  year    = {2022}
}

@article{panaretos2019statistical,
  author    = {Panaretos, Victor M. and Zemel, Yoav},
  title     = {Statistical Aspects of Wasserstein Distances},
  journal   = {Annual Review of Statistics and Its Application},
  year      = {2019},
  volume    = {6},
  number    = {Volume 6, 2019},
  pages     = {405-431},
  doi       = {https://doi.org/10.1146/annurev-statistics-030718-104938},
  publisher = {Annual Reviews},
  issn      = {2326-831X},
  type      = {Journal Article}
}

@article{chavda2014transition,
  author  = {Chavda, N. D. and Deota, H. N. and Kota, V. K. B.},
  title   = {Poisson to GOE transition in the distribution of the ratio of consecutive level spacings},
  journal = {Physics Letters A},
  volume  = {378},
  number  = {41},
  pages   = {3012--3017},
  year    = {2014},
  month   = {8},
  day     = {22}
}

@article{cotler2017chaos,
  author   = {Cotler, Jordan
              and Hunter-Jones, Nicholas
              and Liu, Junyu
              and Yoshida, Beni},
  title    = {Chaos, complexity, and random matrices},
  journal  = {Journal of High Energy Physics},
  year     = {2017},
  month    = {Nov},
  day      = {09},
  volume   = {2017},
  number   = {11},
  pages    = {48},
  issn     = {1029-8479},
  doi      = {10.1007/JHEP11(2017)048}
}

@article{porter1956fluctuations,
  title     = {Fluctuations of Nuclear Reaction Widths},
  author    = {Porter, C. E. and Thomas, R. G.},
  journal   = {Phys. Rev.},
  volume    = {104},
  issue     = {2},
  pages     = {483--491},
  year      = {1956},
  month     = {Oct},
  publisher = {American Physical Society},
  doi       = {10.1103/PhysRev.104.483}
}

@article{curtiss1941on,
  author    = {J. H. Curtiss},
  title     = {{On the Distribution of the Quotient of Two Chance Variables}},
  volume    = {12},
  journal   = {The Annals of Mathematical Statistics},
  number    = {4},
  publisher = {Institute of Mathematical Statistics},
  pages     = {409 -- 421},
  year      = {1941},
  doi       = {10.1214/aoms/1177731679}
}

@article{peropadre2017equivalence,
  title     = {Equivalence between spin Hamiltonians and boson sampling},
  author    = {Peropadre, Borja and Aspuru-Guzik, Al\'an and Garc\'{\i}a-Ripoll, Juan Jos\'e},
  journal   = {Phys. Rev. A},
  volume    = {95},
  issue     = {3},
  pages     = {032327},
  numpages  = {8},
  year      = {2017},
  month     = {Mar},
  publisher = {American Physical Society},
  doi       = {10.1103/PhysRevA.95.032327}
}

@article{wisniacki_PRE2019_OTOC,
  title     = {Gauging classical and quantum integrability through out-of-time-ordered correlators},
  author    = {Fortes, Emiliano M. and Garc\'{\i}a-Mata, Ignacio and Jalabert, Rodolfo A. and Wisniacki, Diego A.},
  journal   = {Phys. Rev. E},
  volume    = {100},
  issue     = {4},
  pages     = {042201},
  numpages  = {13},
  year      = {2019},
  month     = {Oct},
  publisher = {American Physical Society},
  doi       = {10.1103/PhysRevE.100.042201}
}

@article{gisin2002quantum,
  title={Quantum cryptography},
  author={Gisin, Nicolas and Ribordy, Gr{\'e}goire and Tittel, Wolfgang and Zbinden, Hugo},
  journal={Reviews of modern physics},
  volume={74},
  number={1},
  pages={145},
  year={2002},
  publisher={APS}
}

@article{georgescu2014quantum,
  title={Quantum simulation},
  author={Georgescu, Iulia M and Ashhab, Sahel and Nori, Franco},
  journal={Reviews of Modern Physics},
  volume={86},
  number={1},
  pages={153--185},
  year={2014},
  publisher={APS}
}

@article{zhan2026loop,
  title={Loop Quantum Photonic Chip for Coherent Multi-Time-Step Evolution},
  author={Zhan, Yuancheng and Zhang, Hui and Erbanni, Rebecca and Burger, Andreas and Wan, Lingxiao and Jiang, Xudong and Chae, Sanghoon and Liu, Ai Qun and Poletti, Dario and Kwek, Leong Chuan},
  journal={Laser \& Photonics Reviews},
  pages={e02699},
  year={2026},
  publisher={Wiley Online Library}
}

@article{zhang2026integrated,
  title={Integrated Photonic Quantum Computing: From Silicon to Lithium Niobate},
  author={Zhang, Hui and Ma, Yiming and Zhu, Di and Zhan, Yuancheng and Shi, Yuzhi and Wang, Zhanshan and Kwek, Leong Chuan and Laing, Anthony and Liu, Ai Qun and Loncar, Marko and others},
  journal={arXiv preprint arXiv:2601.16484},
  year={2026}
}

@article{bennett2000quantum,
  title={Quantum information and computation},
  author={Bennett, Charles H and DiVincenzo, David P},
  journal={Nature},
  volume={404},
  number={6775},
  pages={247--255},
  year={2000},
  publisher={Nature Publishing Group UK London}
}

@article{pirandola2020advances,
  title={Advances in quantum cryptography},
  author={Pirandola, Stefano and Andersen, Ulrik L and Banchi, Leonardo and Berta, Mario and Bunandar, Darius and Colbeck, Roger and Englund, Dirk and Gehring, Tobias and Lupo, Cosmo and Ottaviani, Carlo and others},
  journal={Advances in optics and photonics},
  volume={12},
  number={4},
  pages={1012--1236},
  year={2020},
  publisher={Optical Society of America}
}

@article{ekert2002direct,
  title={Direct estimations of linear and nonlinear functionals of a quantum state},
  author={Ekert, Artur K and Alves, Carolina Moura and Oi, Daniel KL and Horodecki, Micha{\l} and Horodecki, Pawe{\l} and Kwek, Leong Chuan},
  journal={Physical review letters},
  volume={88},
  number={21},
  pages={217901},
  year={2002},
  publisher={APS}
}

@article{zhang2019integrated,
  title={An integrated silicon photonic chip platform for continuous-variable quantum key distribution},
  author={Zhang, Gong and Haw, Jing Yan and Cai, Hong and Xu, Feng and Assad, SM and Fitzsimons, Joseph F and Zhou, Xianzhong and Zhang, Y and Yu, S and Wu, J and others},
  journal={Nature Photonics},
  volume={13},
  number={12},
  pages={839--842},
  year={2019},
  publisher={Nature Publishing Group UK London}
}

@article{luo2023recent,
  title={Recent progress in quantum photonic chips for quantum communication and internet},
  author={Luo, Wei and Cao, Lin and Shi, Yuzhi and Wan, Lingxiao and Zhang, Hui and Li, Shuyi and Chen, Guanyu and Li, Yuan and Li, Sijin and Wang, Yunxiang and others},
  journal={Light: Science \& Applications},
  volume={12},
  number={1},
  pages={175},
  year={2023},
  publisher={Nature Publishing Group UK London}
}

@article{wang2023experimental,
  title={Experimental boson sampling enabling cryptographic one-way function},
  author={Wang, Xiao-Wei and Zhou, Wen-Hao and Fu, Yu-Xuan and Gao, Jun and Lu, Yong-Heng and Chang, Yi-Jun and Qiao, Lu-Feng and Ren, Ruo-Jing and Jiang, Ze-Kun and Jiao, Zhi-Qiang and others},
  journal={Physical Review Letters},
  volume={130},
  number={6},
  pages={060802},
  year={2023},
  publisher={APS}
}

@article{shannon1948mathematical,
  title={A mathematical theory of communication},
  author={Shannon, Claude Elwood},
  journal={The Bell system technical journal},
  volume={27},
  number={3},
  pages={379--423},
  year={1948},
  publisher={Nokia Bell Labs}
}

@article{alessio2016chaos,
author = {Luca D'Alessio and Yariv Kafri and Anatoli Polkovnikov and Marcos Rigol},
title = {From quantum chaos and eigenstate thermalization to statistical mechanics and thermodynamics},
journal = {Advances in Physics},
volume = {65},
number = {3},
pages = {239--362},
year = {2016},
publisher = {Taylor \& Francis},
doi = {10.1080/00018732.2016.1198134},
eprint = { 
        https://doi.org/10.1080/00018732.2016.1198134
}
}

@article{chertkov2022holographic,
  title={Holographic dynamics simulations with a trapped-ion quantum computer},
  author={Chertkov, Eli and Bohnet, Justin and Francois, David and Gaebler, John and Gresh, Dan and Hankin, Aaron and Lee, Kenny and Hayes, David and Neyenhuis, Brian and Stutz, Russell and others},
  journal={Nature Physics},
  volume={18},
  number={9},
  pages={1074--1079},
  year={2022},
  publisher={Nature Publishing Group UK London}
}

\end{document}